\PassOptionsToPackage{unicode}{hyperref}
\PassOptionsToPackage{hyphens}{url}
\PassOptionsToPackage{dvipsnames,svgnames,x11names}{xcolor}
\documentclass[
  11pt,
  a4paper,
]{article}
\usepackage{xcolor}
\usepackage[margin=2.4cm]{geometry}
\usepackage{amsmath,amssymb}
\usepackage{iftex}
\ifPDFTeX
  \usepackage[T1]{fontenc}
  \usepackage[utf8]{inputenc}
  \usepackage{textcomp} % provide euro and other symbols
\else % if luatex or xetex
  \usepackage{unicode-math} % this also loads fontspec
  \defaultfontfeatures{Scale=MatchLowercase}
  \defaultfontfeatures[\rmfamily]{Ligatures=TeX,Scale=1}
\fi
\usepackage{lmodern}
\ifPDFTeX\else
\fi
\IfFileExists{upquote.sty}{\usepackage{upquote}}{}
\IfFileExists{microtype.sty}{% use microtype if available
  \usepackage[]{microtype}
  \UseMicrotypeSet[protrusion]{basicmath} % disable protrusion for tt fonts
}{}
\makeatletter
\@ifundefined{KOMAClassName}{% if non-KOMA class
  \IfFileExists{parskip.sty}{%
    \usepackage{parskip}
  }{% else
    \setlength{\parindent}{0pt}
    \setlength{\parskip}{6pt plus 2pt minus 1pt}}
}{% if KOMA class
  \KOMAoptions{parskip=half}}
\makeatother
\usepackage{longtable,booktabs,array}
\usepackage{calc} % for calculating minipage widths
\usepackage{etoolbox}
\makeatletter
\patchcmd\longtable{\par}{\if@noskipsec\mbox{}\fi\par}{}{}
\makeatother
\IfFileExists{footnotehyper.sty}{\usepackage{footnotehyper}}{\usepackage{footnote}}
\makesavenoteenv{longtable}
\usepackage{graphicx}
\makeatletter
\newsavebox\pandoc@box
\newcommand*\pandocbounded[1]{% scales image to fit in text height/width
  \sbox\pandoc@box{#1}%
  \Gscale@div\@tempa{\textheight}{\dimexpr\ht\pandoc@box+\dp\pandoc@box\relax}%
  \Gscale@div\@tempb{\linewidth}{\wd\pandoc@box}%
  \ifdim\@tempb\p@<\@tempa\p@\let\@tempa\@tempb\fi% select the smaller of both
  \ifdim\@tempa\p@<\p@\scalebox{\@tempa}{\usebox\pandoc@box}%
  \else\usebox{\pandoc@box}%
  \fi%
}
\def\fps@figure{htbp}
\makeatother
\NewDocumentCommand\citeproctext{}{}
\NewDocumentCommand\citeproc{mm}{%
  \begingroup\def\citeproctext{#2}\cite{#1}\endgroup}
\makeatletter
 \let\@cite@ofmt\@firstofone
 \def\@biblabel#1{}
 \def\@cite#1#2{{#1\if@tempswa , #2\fi}}
\makeatother
\newlength{\cslhangindent}
\newlength{\csllabelwidth}
\newenvironment{CSLReferences}[2] % #1 hanging-indent, #2 entry-spacing
 {\begin{list}{}{%
  \setlength{\itemindent}{0pt}
  \setlength{\leftmargin}{0pt}
  \setlength{\parsep}{0pt}
  \ifodd #1
   \setlength{\leftmargin}{\cslhangindent}
   \setlength{\itemindent}{-1\cslhangindent}
  \fi
  \setlength{\itemsep}{#2\baselineskip}}}
 {\end{list}}
\usepackage{calc}

\providecommand{\tightlist}{%
  \setlength{\itemsep}{0pt}\setlength{\parskip}{0pt}}
\usepackage{amsmath}
\usepackage{amssymb}

\usepackage{newunicodechar}
\AtBeginDocument{%
  \newunicodechar{→}{\ensuremath{\rightarrow}}%
  \newunicodechar{↦}{\ensuremath{\mapsto}}%
  \newunicodechar{↔}{\ensuremath{\leftrightarrow}}%
  \newunicodechar{≈}{\ensuremath{\approx}}%
  \newunicodechar{≤}{\ensuremath{\leq}}%
  \newunicodechar{≥}{\ensuremath{\geq}}%
  \newunicodechar{∈}{\ensuremath{\in}}%
  \newunicodechar{∩}{\ensuremath{\cap}}%
  \newunicodechar{⊂}{\ensuremath{\subset}}%
  \newunicodechar{∞}{\ensuremath{\infty}}%
  \newunicodechar{×}{\ensuremath{\times}}%
  \newunicodechar{±}{\ensuremath{\pm}}%
  \newunicodechar{−}{\ensuremath{-}}%
  \newunicodechar{√}{\ensuremath{\surd}}%
  \newunicodechar{∇}{\ensuremath{\nabla}}%
  \newunicodechar{ℤ}{\ensuremath{\mathbb{Z}}}%
  \newunicodechar{ℝ}{\ensuremath{\mathbb{R}}}%
  \newunicodechar{Ω}{\ensuremath{\Omega}}%
  \newunicodechar{Σ}{\ensuremath{\Sigma}}%
  \newunicodechar{Λ}{\ensuremath{\Lambda}}%
  \newunicodechar{α}{\ensuremath{\alpha}}%
  \newunicodechar{σ}{\ensuremath{\sigma}}%
  \newunicodechar{…}{\ldots}%
  \newunicodechar{′}{\ensuremath{{}^\prime}}%
  \newunicodechar{·}{\textperiodcentered}%
  \newunicodechar{°}{\ensuremath{^\circ}}%
  \newunicodechar{²}{\textsuperscript{2}}%
  \newunicodechar{³}{\textsuperscript{3}}%
  \newunicodechar{¹}{\textsuperscript{1}}%
  \newunicodechar{⁴}{\textsuperscript{4}}%
  \newunicodechar{⁵}{\textsuperscript{5}}%
  \newunicodechar{⁶}{\textsuperscript{6}}%
  \newunicodechar{⁷}{\textsuperscript{7}}%
  \newunicodechar{⁸}{\textsuperscript{8}}%
  \newunicodechar{⁹}{\textsuperscript{9}}%
  \newunicodechar{⁰}{\textsuperscript{0}}%
  \newunicodechar{⁻}{\textsuperscript{\textminus}}%
  \newunicodechar{₀}{\textsubscript{0}}%
  \newunicodechar{₁}{\textsubscript{1}}%
  \newunicodechar{₂}{\textsubscript{2}}%
  \newunicodechar{₃}{\textsubscript{3}}%
  \newunicodechar{₄}{\textsubscript{4}}%
  \newunicodechar{₅}{\textsubscript{5}}%
}

\usepackage{etoolbox}
\AtBeginEnvironment{longtable}{\footnotesize}

\usepackage[htt]{hyphenat}

\makeatletter
\renewcommand\@makefntext[1]{%
  \raggedright\parindent 1em\noindent\hb@xt@1.8em{\hss\@makefnmark}#1}
\makeatother

\usepackage{graphicx}
\graphicspath{{paper_figures/}}

\usepackage{tikz}
\usetikzlibrary{positioning,arrows.meta}

\usepackage{lineno}
\usepackage{bookmark}
\IfFileExists{xurl.sty}{\usepackage{xurl}}{} % add URL line breaks if available
\makeatletter
\@ifundefined{xmpquote}{}{}
\makeatother
\hypersetup{
  pdftitle={Hex9: A Quasi-Authalic{,} Quasi-Continuous Hexagonal DGGS on the Reference Ellipsoid},
  pdfauthor={Ben Griffin},
  colorlinks=true,
  linkcolor={RoyalBlue},
  filecolor={Maroon},
  citecolor={RoyalBlue},
  urlcolor={RoyalBlue},
  pdfcreator={LaTeX via pandoc}}

\title{Hex9: A Quasi-Authalic, Quasi-Continuous Hexagonal DGGS on the
Reference Ellipsoid}
\author{Ben Griffin}
\date{July 2026}

\begin{document}
\maketitle
\begin{abstract}
Discrete global grid systems are conventionally designed over a prior
coordinate reference system and inherit its compromises. Hex9 inverts
the direction of design: we ask what requirements a hierarchical grid
must satisfy to be geometrically coherent --- intrinsic orientability,
flat mode transport, vertex closure, refinement commutativity --- and
show that these requirements, taken together, essentially determine the
grid. The admissible cell primitive is the triangle; the admissible seed
is the octahedral triangulation of S²; admissible refinement is uniquely
aperture 9 --- the only factor whose octant hexagonalises by
half-hexagons, and uniquely up to chirality at that, so the hexagonal
dual lattice admits exactly one orientation per chirality --- both fixed
by the same machine-verified exhaustive enumeration. The structure that
survives is a shifted-aperture-9 hexagonal hierarchy in which every cell
carries a unique address derived from the construction alone: truncated
at level L, the address is a DGGS zonal identifier in the sense of OGC
Topic 21; carried to the limit, a function recovers from it a point on
the reference ellipsoid to arbitrary precision. The addressing is
quasi-continuous --- position is recoverable everywhere except on a
measure-zero set of seams --- rather than continuous in the strict ISO
19111 sense; the same mathematical object serves as both zonal
identifier and position-recovery coordinate, with no prior coordinate
reference system as input. A separable geometric realisation --- an
analytical octahedral base projection composed with an
optimal-transport-derived area-correcting warp --- places the grid on
WGS84 with quasi-uniform cell areas: at level 5, 99\% of the 708,588
cells lie within 0.005\% of ideal area, with residual deviation confined
to the six octahedral vertices required by the topology. The
combinatorial grid is projection- and ellipsoid-independent; only the
warp is specific to the reference body, and it is recomputable for any
ellipsoid, terrestrial or planetary.

\end{abstract}

{
\setcounter{tocdepth}{2}
\tableofcontents
}
\subsection{Introduction}\label{introduction}

Discrete global grid systems face a set of familiar tradeoffs. Cell
shape, aperture, orientation, and global embedding each involve
compromises between equal area, hierarchical consistency, polar
behaviour, and computational tractability --- compromises often governed
by a prior choice of coordinate reference system, which in turn
constrains what tradeoffs remain.

\begin{figure}
\centering
\includegraphics[width=1\linewidth,height=\textheight,keepaspectratio,alt={The Hex9 level-2 grid (972 cells) over NASA Blue Marble imagery, in the system's native space: the b\_oct octahedral ``butterfly'' net (§11). In b\_oct every cell is a congruent regular hexagon (§11c) --- the reshaping of the familiar continents is the work of the octahedral unfolding, not of the cells. Conventional reprojected views of the same grid appear in §14c.}]{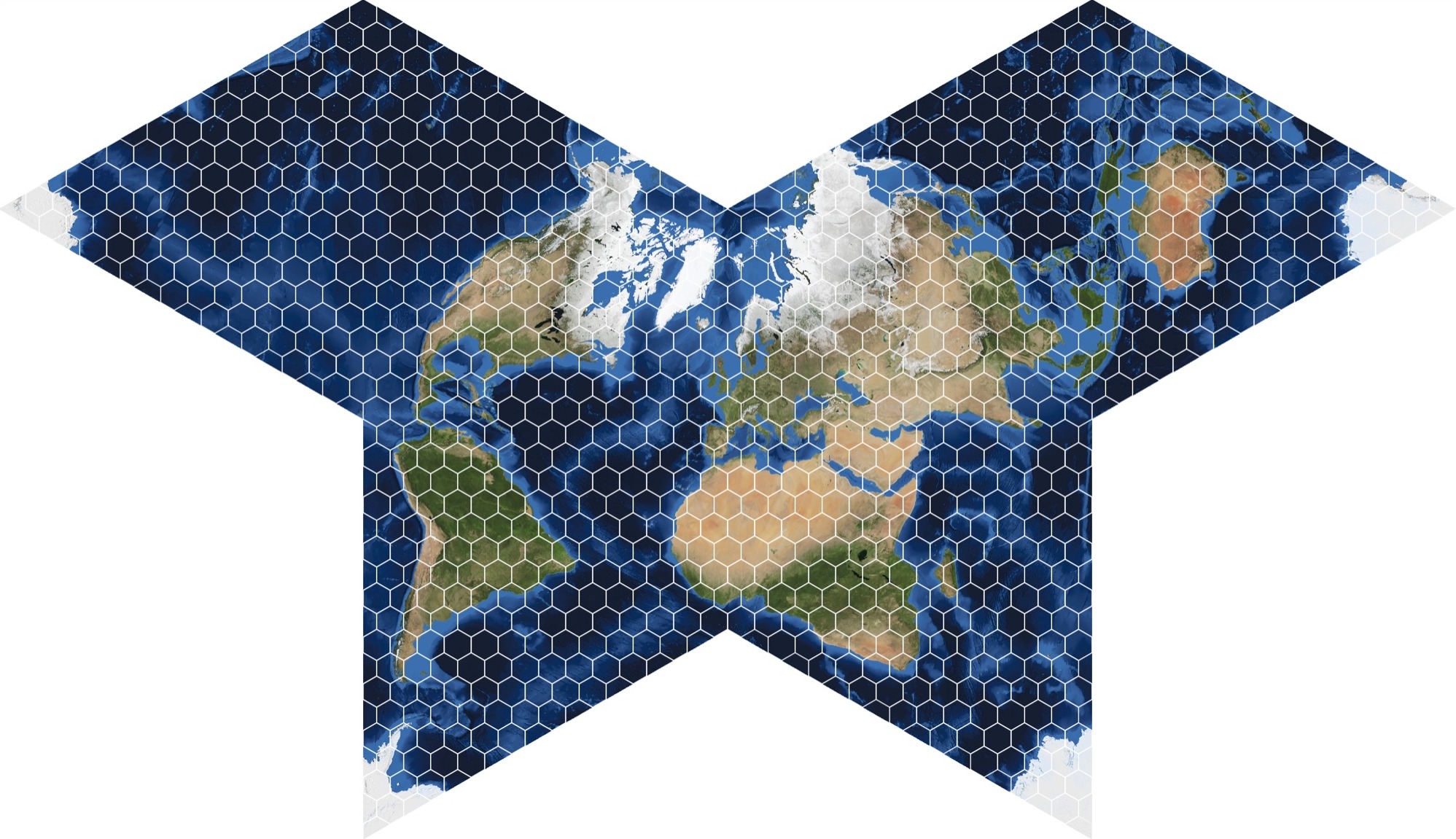}
\caption[The Hex9 level-2 grid (972 cells) over NASA Blue Marble
imagery, in the system's native space: the b\_oct octahedral
``butterfly'' net (§11). In b\_oct every cell is a congruent regular
hexagon (§11c) --- the reshaping of the familiar continents is the work
of the octahedral unfolding, not of the cells. Conventional reprojected
views of the same grid appear in §14c.]{The Hex9 level-2 grid (972
cells) over NASA Blue Marble imagery, in the system's native space: the
b\_oct octahedral ``butterfly'' net (§11). In b\_oct every cell is a
congruent regular hexagon (§11c) --- the reshaping of the familiar
continents is the work of the octahedral unfolding, not of the cells.
Conventional reprojected views of the same grid appear in
§14c.\footnotemark{}}
\end{figure}
\footnotetext{Generated by
  \protect\texttt{examples/ex0097\_smp\_grid.py}
  (\protect\texttt{butterfly:0500}, L2).}

Hex9 is an exploration of a different resolution strategy: rather than
choosing among tradeoffs, we ask what requirements a grid must satisfy
to be geometrically coherent --- consistent at every edge, every vertex,
and every level of the hierarchy --- and examine how far those
requirements determine the structure on their own.

These requirements significantly reduce the design freedom. A minimal
set --- orientability, parity transport, vertex closure, and refinement
commutativity --- strongly constrains admissible cell shape, embedding
class, and aperture structure. What appears to be a large design space
resolves to a small family of solutions, selecting a restricted class up
to global symmetry.

This derivation has a direct consequence: because the structure is
derived rather than chosen, every cell carries a distinct identity tied
to a specific surface location. Cell identity serves as a self-contained
locator --- no prior coordinate reference system is needed to establish
where a cell is.

The derivation was discovered rather than designed. The project began as
a planar hexagonal tiling study and first targeted the icosahedron ---
the natural choice, given its near-spherical geometry and the precedent
of existing icosahedral grids. The tiling could not be made globally
consistent, and the failure resisted diagnosis: the obstruction is not
geometric but topological. The odd valence of icosahedral vertices makes
a consistent two-colouring of faces impossible, and that two-colouring
turns out to be load-bearing for everything a coherent hexagonal
hierarchy needs. Once the parity constraint was understood, the
octahedron emerged by elimination as the unique Platonic solid with
equilateral faces and even-valent vertices. The system presented here is
the consequence of following that constraint to its conclusions.

The paper develops this argument constructively. The first part traces
the coherence arc, showing step by step how each requirement eliminates
alternatives. The second part describes the AK+Warp projection ---
Anders Kaseorg's octahedral base projection composed with an
area-correcting warp --- the engineering path from the abstract
octahedral structure to a quasi-authalic realisation on the reference
ellipsoid (typically WGS84).

\begin{center}\rule{0.5\linewidth}{0.5pt}\end{center}

\subsection{Axiom Set}\label{axiom-set}

\subsubsection{Axiom 1 --- Domain (Ellipsoidal
Manifold)}\label{axiom-1-domain-ellipsoidal-manifold}

The Earth is modelled as a smooth reference ellipsoid (e.g.~WGS84),
supporting a well-defined normal field and latitude/longitude
parameterisation.

\begin{itemize}
\tightlist
\item
  This is the continuous substrate being discretised.
\item
  No projection is privileged at this level.
\end{itemize}

\subsubsection{Axiom 2 --- Discrete Carrier (Simplicial
Primacy)}\label{axiom-2-discrete-carrier-simplicial-primacy}

The ellipsoid is discretised by a finite, recursively refinable 2D
simplicial complex.

\begin{itemize}
\tightlist
\item
  The primitive cell is a triangle (2-simplex).
\item
  All higher structure is derived from simplicial adjacency.
\item
  No non-simplicial base cells exist in the canonical representation.
\end{itemize}

\emph{Consequence: hex-like structures are derived, not primitive.}

\subsubsection{Axiom 3 --- Global Topological
Regularity}\label{axiom-3-global-topological-regularity}

The simplicial complex forms a valid triangulation of S² with:

\begin{itemize}
\tightlist
\item
  no boundary,
\item
  no exceptional cell types,
\item
  curvature expressed metrically, not topologically.
\end{itemize}

This enforces uniform representation class across the entire domain and
excludes complexes with topological defects (such as forced pentagonal
faces).

\emph{Consequence: combined with Axiom 2 and Axiom 4 --- simplicial
carrier, even valence from vertex closure, and this axiom's uniformity
requirement --- the admissible complex is the octahedral triangulation.}

\subsubsection{Axiom 4 --- Mode Transport}\label{axiom-4-mode-transport}

Mode transport is consistent when traversing any closed loop through the
triangulation returns the mode to its starting value (trivial holonomy,
in the language of differential geometry; equivalently, a flat ℤ₂
connection, in gauge-theoretic terms). This is equivalent to the
existence of a globally coherent mode field: fix the mode of any one
face and every other face's mode follows directly.

Any refinement operator must preserve this consistency. Failure to do so
introduces a transport defect.

\emph{Consequence (proved in §6): a refinement hexagonalises only if its
octant tiles by half-hexagons, and aperture 9 (linear factor 3) is the
unique factor that does so uniquely up to chirality --- apertures 4, 16,
25, 49 cannot tile at all; 36, 81, \ldots{} tile in many inequivalent
ways. Aperture 9 is therefore not merely minimal but the only refinement
that determines a grid.}

\subsubsection{Axiom 5 --- Refinement
Invariance}\label{axiom-5-refinement-invariance}

Any refinement operator:

\begin{itemize}
\tightlist
\item
  preserves simplicial type,
\item
  preserves adjacency relations,
\item
  commutes with global indexing,
\item
  preserves parity transport (Axiom 4).
\end{itemize}

Refinement never introduces new structural categories.

\subsubsection{Axiom 6 --- Canonical Orientation (Chirality
Fixing)}\label{axiom-6-canonical-orientation-chirality-fixing}

The global orientation of simplices is fixed by a single consistent
chirality choice induced by embedding into the reference ellipsoid
frame.

\begin{itemize}
\tightlist
\item
  This resolves the residual Z₂ symmetry left after Axioms 2--5.
\item
  The choice is globally coherent, not locally independent.
\end{itemize}

\emph{Consequence: this is where the half-hexagon orientation freedom
collapses to one surviving solution per chirality.}

\subsubsection{Axiom 7 --- Geodetic Anchoring (Minimal Frame
Fixing)}\label{axiom-7-geodetic-anchoring-minimal-frame-fixing}

A minimal geodetic frame is fixed:

\begin{itemize}
\tightlist
\item
  poles define the primary axis,
\item
  a single meridian defines longitudinal zero,
\item
  together inducing a canonical partition of the domain.
\end{itemize}

This is not arbitrary once fixed --- it is the reference gauge of the
system.

\subsubsection{Axiom 8 --- Unique Cell
Addressing}\label{axiom-8-unique-cell-addressing}

Every cell in the hierarchy has a unique address, and every valid
address identifies exactly one cell. This bijection is compatible with
hierarchical structure: a cell's address encodes its position in the
refinement hierarchy, and the parent--child relationship is recoverable
directly from address structure.

Combinatorial position is read directly from the address: cell identity
and the parent--child hierarchy require no transformation. Geographic
location follows from that position through the system's own canonical
realisation (§11), a fixed map that adds no degrees of freedom --- so no
prior or external coordinate reference system is required as input. The
grid fixes the discrete structure; the realisation supplies the metric
embedding; the two are separable concerns (§11) composed to recover a
location from an address.

\subsubsection{Axiom 9 --- Dual Consistency (Hex
Emergence)}\label{axiom-9-dual-consistency-hex-emergence}

The dual graph of the simplicial complex induces a structured hexagonal
lattice:

\begin{itemize}
\tightlist
\item
  hexagons are derived dual cells,
\item
  the total topological defect of 12 required by the Euler
  characteristic of S² (Σ over vertices of (6 − valence) = 12) is
  concentrated at the 6 seed vertices --- a deficit of 2 at each ---
  rather than appearing as face anomalies,
\item
  no independent hex tiling is assumed at the primitive level.
\end{itemize}

\emph{Hex9 structure is emergent from simplices, not fundamental.}

\subsubsection{Main Theorem}\label{main-theorem}

Axioms 1--9 admit a discrete global grid that is \textbf{unique up to
two free choices} --- a global chirality (Axiom 6) and a geodetic gauge
(Axiom 7). That grid is \textbf{Hex9}: the shifted-aperture-9 hexagonal
grid on the octahedral triangulation of the reference ellipsoid, with
the cell hierarchy and addressing developed in §§6--10.

Concretely, the axioms force each link of the following chain, and the
arc sections that follow constitute its proof:

\begin{enumerate}
\def\labelenumi{\arabic{enumi}.}
\tightlist
\item
  \textbf{Carrier.} The discrete carrier is a simplicial complex, and
  the only triangulation of S² with even, uniform valence and no
  exceptional cells is the octahedral triangulation (Axioms 2--4;
  §§1--5).
\item
  \textbf{Aperture.} A hexagonalisation requires the octant to tile by
  half-hexagons, and \textbf{only aperture 9 (linear factor 3) does so
  uniquely up to chirality} --- the same exhaustive enumeration that
  fixes the chiral pair (Axiom 6). Apertures 4, 16, 25, 49 cannot tile
  at all; 36, 81, \ldots{} tile in many inequivalent ways. Aperture 9 is
  the unique refinement that determines a grid (Axioms 4--6; §6).
\item
  \textbf{Hex emergence.} The dual carries the 12 units of topological
  defect required by the Euler characteristic of S² as twelve
  five-neighbour hexagons at the six octahedral vertices, not as a
  distinct cell type (Axiom 9; §7).
\item
  \textbf{Orientation.} The residual orientation freedom collapses to a
  single chiral pair (Axiom 6; §8).
\item
  \textbf{Identity.} Every cell carries a unique hierarchical address
  from which its geographic location is read directly, with no prior
  coordinate reference system (Axiom 8; §§9--10).
\end{enumerate}

The geometric realisation on the ellipsoid (§11) is a separable
engineering step: it places this combinatorial object onto WGS84 but
adds no degrees of freedom to the structure the axioms determine.

\begin{center}\rule{0.5\linewidth}{0.5pt}\end{center}

\subsection{1. The Simplicial Carrier}\label{the-simplicial-carrier}

A discrete global grid partitions the globe into a finite collection of
cells. The choice of cell shape is the first apparent design decision:
squares, triangles, hexagons, and other polygons all tile the plane, and
several tile the globe with appropriate modifications.

The triangle stands apart from this class. It is the minimal polygon ---
three edges, three vertices, no simpler closed shape exists --- and the
only one whose orientation is intrinsic: given any triangle, a
consistent notion of clockwise and counterclockwise is determined by its
vertex ordering alone, without additional structure. Every other polygon
either decomposes into triangles or requires external reference to
define orientation consistently across a tiling.

This intrinsic orientability makes the triangle the natural substrate
for a globally coherent grid. We take the triangular field --- a tiling
of the globe by triangles --- as the simplicial carrier on which the
remaining coherence requirements will act.

A coherent grid carrier must satisfy several requirements
simultaneously: operators such as interpolation and refinement must be
definable without auxiliary choices; each cell must support a unique,
intrinsic coordinate system; refinement must be closed and
structure-preserving; and no cell-type exceptions or singular vertices
may appear. The triangle is the only 2-cell that satisfies all of these
at once. Its barycentric coordinates are intrinsic --- defined by the
vertices alone, with no ambient metric required. It is closed under
subdivision. It forms a simplicial complex without diagonal ambiguity.

On curved surfaces this distinction becomes decisive. A triangulation
absorbs curvature at its vertices without changing cell type; tilings by
higher polygons must place their topological obligations somewhere
visible --- as exceptional cells (the twelve pentagons of icosahedral
hexagonal grids) or exceptional corners (the eight 3-valent corners of
cube-based quadrilateral grids). The triangular field is therefore not
one choice among many --- every other option either reduces to it or
requires additional structure not established by the coherence
requirements alone.

\begin{center}\rule{0.5\linewidth}{0.5pt}\end{center}

\subsection{2. Mode}\label{mode}

In any consistent triangulation of an orientable surface, every edge is
shared by exactly two triangles. The intrinsic orientation established
in §1 ensures those two triangles carry opposite orientations: no two
triangles of the same orientation can share an edge.

The result is a consistent parity assignment over faces induced by
orientation. In the familiar planar picture these are the \emph{up} and
\emph{down} triangles; on the globe they generalise to two classes that
partition every face in the triangulation. We call this two-colouring
the \textbf{mode} of the triangular field --- mode 0 for the down
(negative) class, mode 1 for the up (positive) class. This labelling is
not imposed from outside --- it is a structural property of the
triangulation itself. Fix the mode of any one face, and the mode of
every other face follows directly; the only freedom is which face to
start from, which amounts to a global reflection.

Every face carries a mode value. Every interior edge is incident to
exactly two faces of opposite mode. The field is bipartite on its faces.

As the ℤ₂ orientation cocycle of the simplicial surface, realised as a
global face bipartition, \textbf{mode} is the first emergent global
invariant of simplicial orientation --- a constraint every grid built
over the field must either respect or violate. Everything downstream
either respects that cocycle, preserving mode consistently across
refinement and adjacency, or breaks it, introducing inconsistency or
defects. Every edge crossing is a mode-flipping step; what follows from
global consistency of those steps is the subject of §3.

\begin{center}\rule{0.5\linewidth}{0.5pt}\end{center}

\subsection{3. Mode Transport}\label{mode-transport}

Section 2 established that every edge crossing carries a mode flip. The
transport operator on each edge is that flip --- the orientation
difference between the two faces sharing the edge, inherited directly
from the intrinsic orientability of §1. It is combinatorial in origin:
no metric or embedding is required to define it.

We now ask what it means for this transport to be globally consistent.

Define a \textbf{mode transport} as a rule that assigns, to any path
through the triangular field (a sequence of edge crossings from face to
face), a net mode change in ℤ₂. The transport is \textbf{consistent} if
the net change depends only on the endpoints of the path --- not on the
route taken; equivalently, traversing any closed loop returns the mode
to its starting value. This is the condition differential geometry calls
trivial holonomy and gauge theory calls a flat ℤ₂ connection; we will
simply say the transport is \textbf{flat}. On any orientable surface
flatness is equivalent to the existence of a globally consistent mode
field: fix the mode of any one face and every other face's mode follows.
Any refinement must preserve flatness; otherwise it introduces a
transport defect.

The requirement becomes non-trivial when we demand that consistency
extends to all refinements of the triangulation. A refinement introduces
new faces, edges, and vertices --- and the transport must remain
consistent at every scale. This is the content of Axiom 4, and its
consequences for the admissible aperture class are developed in §6.

What the transport implies immediately is simpler: any structure built
over the triangular field --- any labelling, orientation, or subdivision
--- must account for the mode flip at every edge crossing, or introduce
a defect. The mode is not merely a local colouring; it is a global
invariant that any grid must either propagate correctly or break.

At individual edges the transport is binary and clean. At vertices ---
where multiple edges meet --- the accumulated transports must also close
consistently. What this implies for the global topology is the subject
of §4.

\begin{center}\rule{0.5\linewidth}{0.5pt}\end{center}

\subsection{4. Vertex Closure}\label{vertex-closure}

Before the argument, its conclusion: we do not pick the octahedron from
a menu of base shapes. It is the only one on which a true hexagonal grid
can both wrap a closed surface and unfold flat to the plane without
contradiction. What can read as a self-imposed constraint is really the
absence of alternatives --- the rounder shapes other grids build on do
not support the same coherent hexagonal hierarchy.

Section 3 established that mode transport must be flat --- any closed
loop must return the mode to its starting value. Vertices are where this
condition is most constraining: at every vertex, a ring of triangular
faces meets, and traversing that ring forms a closed loop.

Consider a vertex where k triangular faces meet (valence k). Moving from
face to face around the vertex traverses exactly k edges. Each edge
crossing carries a mode flip. For the transport around this closed loop
to return to identity, the total number of flips must be even ---
requiring k to be even.

This is the \textbf{vertex closure condition}: it selects triangulations
in which every vertex has even valence. It ensures that the face
adjacency graph is bipartite --- no odd cycles in the dual.

Face bipartiteness alone does not require uniform valence. Many
irregular triangulations of S² admit a consistent mode assignment while
mixing different even valences. The vertex closure condition, taken
alone, is compatible with non-uniform even valence, and we make no claim
that such triangulations fail under refinement.

The selection of uniform valence comes instead from the regularity
requirement of Axiom 3: the seed complex admits no exceptional cell
types and no distinguished vertex classes. A triangulation mixing
different even valences contains several distinct vertex-star types, and
every structure built over it --- refinement rules, addressing,
adjacency --- would have to distinguish those classes explicitly. Axiom
3 excludes this by requirement, not by theorem: every vertex star is of
the same type, so every vertex behaves identically under refinement and
addressing. Uniform even valence is therefore imposed as a regularity
condition, not derived as a topological necessity.

Given uniform even valence, the Euler characteristic of S² constrains
what is possible. For a triangulation with V vertices, E edges, F faces:

\begin{verbatim}
V − E + F = 2,   with   E = 3F/2
\end{verbatim}

For uniform valence v the relation 2E = vV gives:

\begin{verbatim}
V = 12 / (6 − v)
\end{verbatim}

For v = 4 (the minimum even valence greater than 2): V = 6, F = 8, E =
12. For v = 6: the denominator vanishes --- uniform valence 6 cannot
close on the globe. For v ≥ 8: the formula yields a negative vertex
count --- impossible.

The octahedral triangulation --- V=6, E=12, F=8, uniform valence 4 ---
is the one abstract triangulation of S² satisfying all conditions. The
division of labour is explicit: vertex closure, a derived condition,
forces even valence; regularity (Axiom 3), a stated requirement, forces
uniform valence; the Euler characteristic then leaves v = 4 as the only
possibility.

\emph{Consequence: a ℤ₂ face-mode can be defined on any triangulation of
S² with a bipartite dual graph --- equivalently, on any triangulation in
which every vertex has even valence. Vertex closure is the derived part
of this argument; uniformity of valence is required by Axiom 3 rather
than derived. Together they admit exactly one triangulation of S²: the
octahedral one.}

\begin{center}\rule{0.5\linewidth}{0.5pt}\end{center}

\subsection{5. The Octahedral Embedding}\label{the-octahedral-embedding}

Section 4 established that the minimal regular seed complex compatible
with mode transport on the globe has exactly 6 vertices, 12 edges, and 8
triangular faces, with uniform vertex valence 4. There is exactly one
abstract triangulation of S² with this combinatorial structure --- and
it is the octahedral triangulation. Its geometric realisation as a
convex polyhedron is the regular octahedron.

On the sphere, every embedding of the octahedral seed is equivalent: the
sphere's symmetry group O(3) contains the full octahedral group O\_h in
any orientation, and no embedding is preferred. The reference ellipsoid
breaks this. An ellipsoid of revolution retains only the symmetries
fixing its polar axis --- continuous rotation about the axis, the
equatorial mirror, and the vertical mirror planes (the group D∞h). The
poles are the fixed points of this residual symmetry.

An embedded octahedron shares with the ellipsoid exactly those
symmetries common to both, and the shared group depends on which
octahedral axis is aligned with the polar axis. Aligning a vertex pair
(a 4-fold axis) retains D4h, of order 16; a face pair (a 3-fold axis)
retains D3d, of order 12; an edge pair retains D2h, of order 8; a
generic orientation retains almost nothing. Pole-on-vertex anchoring is
therefore not one choice among equals: it is the unique orientation
class preserving the maximal common symmetry of seed and surface.

This maximal residual symmetry is what the construction uses. Under D4h
the 8 octant faces form a single orbit --- every octant face is
equivalent to every other --- so a single projection function serves all
8 octants, with mode-1 faces obtained from mode-0 by one reflection (y →
−y). The anchoring also reduces the continuous rotational gauge freedom
to the discrete 4-fold rotation, which the meridian anchoring of Axiom 7
resolves. Pole anchoring does not impose a coordinate system; it selects
the embedding in which the seed inherits the most structure from the
surface.

The result is a canonical partition of the globe into 8 triangular
octants --- not a projection choice, but the direct consequence of mode
transport closure applied to a closed orientable surface. The 8 octants
are the top-level cells of a grid hierarchy following these constraints.
Further structure --- refinement, orientation, dual cells --- is built
over this seed. The octahedral embedding is not one possible global
frame among many; it is the frame the coherence requirements construct.

\textbf{Is this a loss of flexibility?} It can read as one --- why
commit to a single base polyhedron when others give rounder cells? ---
but the flexibility is largely illusory. No hexagonal grid tiles a
sphere at all without defects (Euler), and demanding one that
\emph{both} closes on a folded polyhedron \emph{and} unfolds coherently
to the plane, carrying the same two-colouring through every refinement,
is a genuinely narrow requirement: §§4--5 show the octahedron is the
seed that survives, not one rounder option among many that we set aside.
What the commitment then buys is the coherence the rest of the paper
rests on --- the two-colouring (mode transport) is what produces the
half-hexagon hierarchy and its exact, self-inverting nesting (§7, §10e,
§13b), the property §12 sets against iterated hexagonal indices ---
together with an economy of realisation: the eight octant faces are all
equivalent by symmetry, so a single projection and warp serve the whole
globe (§11), and the unavoidable defect sits at six vertices rather than
the twelve of an icosahedral grid. The one real cost is cell shape ---
an octahedral seed is less round per cell than icosahedral or
dodecahedral grids (aspect 1.37 vs H3's 1.06; §12) --- but those rounder
grids do not deliver the same coherent, exactly-nested hexagonal
hierarchy. The choice is between different objects, not a menu of
equivalents.

\begin{center}\rule{0.5\linewidth}{0.5pt}\end{center}

\subsection{6. Refinement Commutativity}\label{refinement-commutativity}

The octahedral seed established in §5 defines a base simplicial complex
on which a hierarchical refinement operator acts. Each triangular face
is subdivided into k² child triangles by linear refinement at scale
factor k. The refinement operator must preserve simplicial type,
adjacency relations, and the flatness of mode transport (§3), and must
commute with the global indexing of the mesh.

Refinement is chosen to commute with mode transport because mode is the
global coherence field of the system. Without commutativity, refinement
would not preserve identity under scale, and hierarchical addressing
would cease to be stable under composition. This choice is not
geometrically mandatory, but it is required for refinement to function
as a consistent coordinate extension rather than a sequence of unrelated
discretisations.

These requirements constrain admissible values of k. First, k must be a
positive integer to ensure that refinement is a well-defined subdivision
of the simplicial structure into congruent refinement classes.

Second, refinements that preserve mode transport consistency are those
in which the mode of every child triangle agrees with the mode inherited
from its parent at every induced edge (a homomorphism of the ℤ₂
transport structure, in algebraic terms).

The mode two-colouring alone does not single out a factor: a parent's
children carry both modes at every scale, for every k, so mode transport
is consistently inherited regardless. What singles out the aperture is a
stronger requirement --- that the refinement actually
\textbf{hexagonalise}, and do so unambiguously.

Two conditions settle it, and a single exhaustive enumeration supplies
both (verified by \texttt{experimental/halfhex\_verify.py}, swept over k
by \texttt{experimental/aperture\_tiling.py}). First, the subdivided
octant --- a side-k triangle of k² triangular cells --- must tile by
half-hexagons (the d\_cells of §7), each three cells, which is possible
only when 3 divides k. Second, and decisively, that tiling must be
\textbf{unique up to chirality}; otherwise the refinement admits several
inequivalent grids and addressing is not stable under composition.

The enumeration is sharp. Apertures 4, 16, 25 and 49 (factors 2, 4, 5,
7) cannot tile the octant by half-hexagons at all. Apertures 36, 81 and
larger \emph{can} tile, but in many inequivalent ways --- 220 distinct
tilings already at aperture 36 --- fixing no canonical grid.
\textbf{Aperture 9 alone (factor 3) tiles uniquely up to chirality:
exactly two tilings, a single mirror pair.} This is not a separate fact
from the chirality result --- it is the same enumeration: the two
tilings \emph{are} the chiral pair that Axiom 6 resolves (revisited in
§8).

Aperture 9 is therefore not merely the smallest admissible refinement;
it is the only one that determines a grid. We take k = 3 as the base
refinement operator, and compose it for finer scales, yielding the
aperture-9 hierarchy.

In the k=3 case, refinement introduces a systematic lateral displacement
of child simplex centroids relative to the parent geometry. This
displacement is not an artefact of embedding, but the geometric
manifestation of parity-preserving refinement under ℤ₂ transport. In the
dual structure, this offset becomes visible as the characteristic shift
in the induced hexagonal lattice, giving rise to the
\textbf{shifted-aperture-9 hierarchy}. Intuitively, the parent's
centroid falls on the shared long edge of its two half-hexagons rather
than on any one child, so the nine children cannot sit concentrically;
they take the offset arrangement of §8, and that offset is the
displacement.

\begin{center}\rule{0.5\linewidth}{0.5pt}\end{center}

\subsection{7. Dual Projection --- Hexagonal
Structure}\label{dual-projection-hexagonal-structure}

The dual of a triangulation exchanges faces and vertices: each
triangular face becomes a dual node, and each vertex becomes a dual face
whose valence equals the vertex degree.

In the refined octahedral triangulation, interior vertices attain
valence 6, producing regular hexagonal dual cells. The six vertices of
the octahedral seed constitute the only non-uniform elements of the
triangulation; each is 4-valent, a vertex deficit of 2, together
accounting for the total defect of 12 required by the Euler
characteristic (§4). The dual cells remain hexagons throughout: the
deficit is carried not by a different cell shape but by twelve hexagons
--- two meeting at each seed vertex --- that there have only five
distinct neighbours, two of their sides folding together across the
vertex. Refinement spreads regular valence-6 structure throughout the
interior while localising this irregularity to the six seed vertices.

It follows that the hexagonal lattice is not imposed but emerges as the
dual of the locally regular simplicial field. The octahedral embedding
provides the minimal finite source of irregularity required by the Euler
characteristic of S², while preserving maximal hexagonal regularity
elsewhere.

Within each octant, the dual lattice forms a half-hexagonal region (the
\textbf{half-hexagon}) bounded by seed edges. The k=3 refinement
introduces a systematic lateral displacement of these regions in the
dual lattice. This shift is the spatial manifestation of enforcing
refinement as a homomorphism of the mode transport structure.

The planar structure has a crystallographic signature. The hexagonal
tiling cut along the half-hexagon long edges in the Hex9 arrangement has
wallpaper group p31m --- a strict reduction from the p6m of the uncut
tiling. Colouring by mode reduces it further: every mirror of p31m
exchanges the two modes (a mode-preserving mirror would make the tiling
achiral, contradicting the chiral pair of §8), so the mode-preserving
subgroup is exactly the rotational part, p3. The ℤ₂ of §2 is, in
crystallographic terms, the quotient p31m / p3.

\begin{figure}
\centering
\includegraphics[width=0.85\linewidth,height=\textheight,keepaspectratio,alt={Crystallographic structure of the d\_cell tiling: the translational unit (left) and its IUC symmetry --- three-fold centres, mirror and glide lines, fundamental domain (right). The symmetry-reduction chain is p6m → p31m → p3 (§7).}]{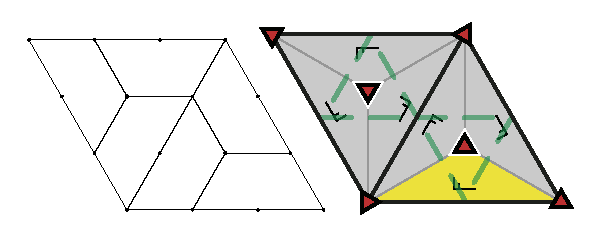}
\caption{Crystallographic structure of the d\_cell tiling: the
translational unit (left) and its IUC symmetry --- three-fold centres,
mirror and glide lines, fundamental domain (right). The
symmetry-reduction chain is p6m → p31m → p3 (§7).}
\end{figure}

The resulting system consists of a hexagonal dual lattice with a
minimal, seed-localised defect structure and a refinement-induced shift.
The remaining degree of freedom is the global and inter-octant
orientation of this lattice.

\begin{center}\rule{0.5\linewidth}{0.5pt}\end{center}

\subsection{8. Orientation Selection}\label{orientation-selection}

The half-hexagon established in §7 is the natural octant carrier --- the
region of the dual hexagonal lattice bounded by the three seed edges of
a single octant face. Its internal structure admits a further
decomposition: the half-hexagon divides into three equilateral
sub-regions, each corresponding to one third of the octant face.

Each sub-region can be independently oriented in two ways: the hexagonal
cells within it may be arranged in one of two configurations related by
reflection. With three sub-regions and two choices each, there are 2³ =
8 candidate orientations for the half-hexagon as a whole.

The mode transport constraint --- specifically, the vertex closure
condition established in §4 --- acts on the internal boundaries between
sub-regions. At every vertex where two or more sub-regions meet, the
mode transport around that vertex must return to identity: trivial
holonomy. This is the same condition that selects for even valence
globally; applied to the internal half-hexagon boundaries, it filters
the 8 candidates.

Exactly 2 of the 8 combinations satisfy the internal closure condition
at every boundary vertex. The two survivors are related by a global
reflection --- a chiral pair, geometrically distinct but structurally
equivalent up to handedness.

\begin{figure}
\centering
\includegraphics[width=1\linewidth,height=\textheight,keepaspectratio,alt={The surviving chiral pair: the two orientations of the 9-cell equilateral that satisfy internal closure (three half-hexagons each, coloured by orientation class). The members are mirror images --- the residual chirality Axiom 6 fixes.}]{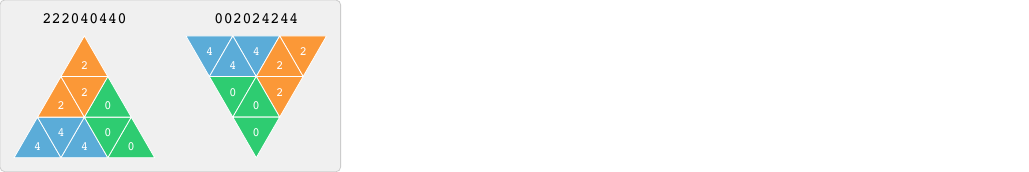}
\caption[The surviving chiral pair: the two orientations of the 9-cell
equilateral that satisfy internal closure (three half-hexagons each,
coloured by orientation class). The members are mirror images --- the
residual chirality Axiom 6 fixes.]{The surviving chiral pair: the two
orientations of the 9-cell equilateral that satisfy internal closure
(three half-hexagons each, coloured by orientation class). The members
are mirror images --- the residual chirality Axiom 6
fixes.\footnotemark{}}
\end{figure}
\footnotetext{Generated by
  \protect\texttt{experimental/halfhex\_further.py}; enumeration counts
  verified by \protect\texttt{experimental/halfhex\_verify.py}.}

This count is not an assertion. The full enumeration is machine-verified
by \texttt{experimental/halfhex\_verify.py} (checks V0--V6): of the 49
distinct hextile solutions (24 chiral pairs + 1 self-mirror), the
long-edge constraint (A) admits 18, the three-equilateral structural
constraint (B) admits 8, and their intersection A ∩ B is exactly the
recorded Hex9 chiral pair. The constructive 2³ → 2 argument above and
this enumeration agree.

\begin{figure}
\centering
\includegraphics[width=0.85\linewidth,height=\textheight,keepaspectratio,alt={The 49 distinct hextile solutions, chiral pairs grouped and the self-mirror marked, with the Hex9 pair highlighted. Of 49 = 24 pairs + 1 self-mirror, constraint A admits 18, constraint B admits 8, and A ∩ B is the highlighted pair.}]{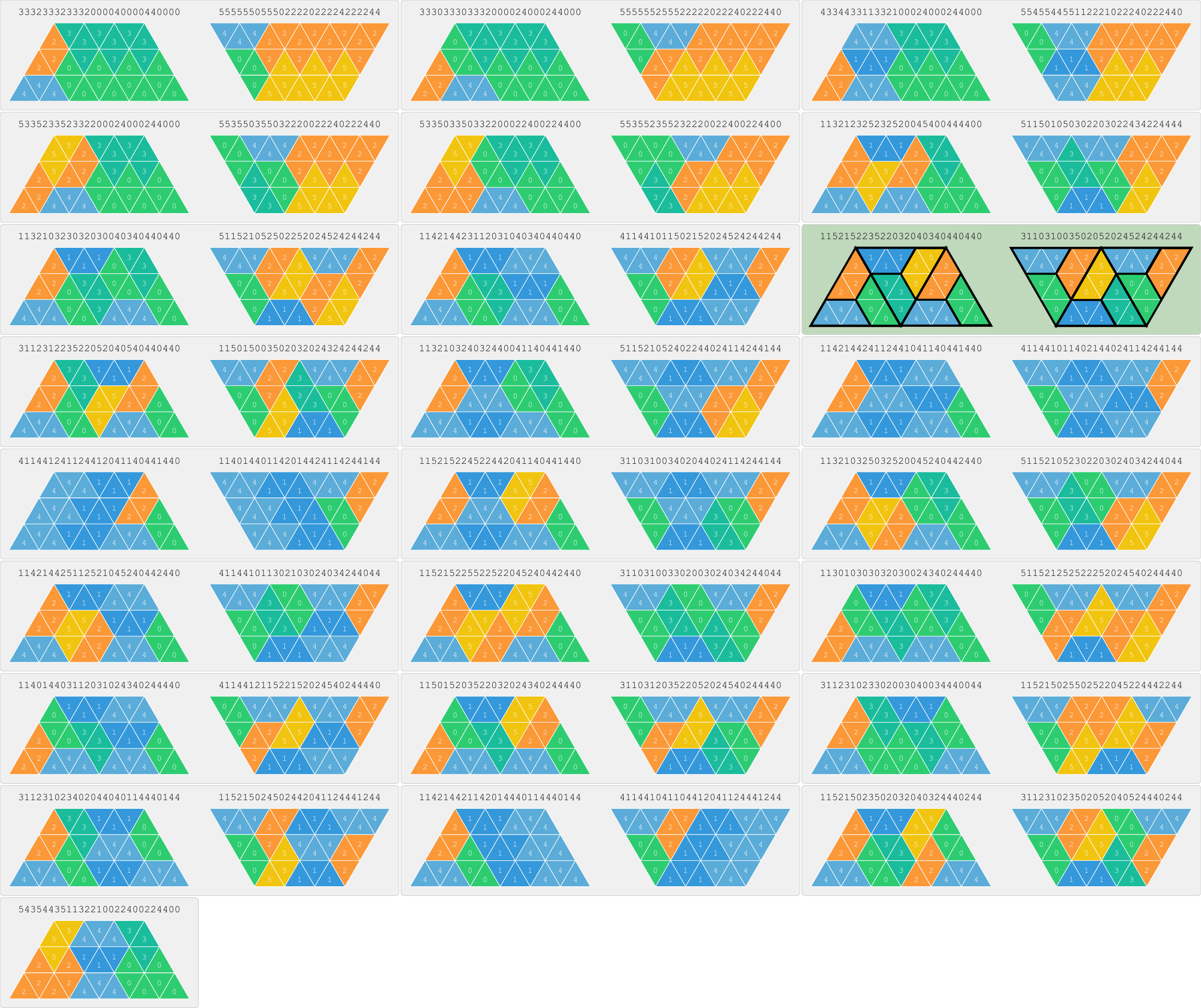}
\caption[The 49 distinct hextile solutions, chiral pairs grouped and the
self-mirror marked, with the Hex9 pair highlighted. Of 49 = 24 pairs + 1
self-mirror, constraint A admits 18, constraint B admits 8, and A ∩ B is
the highlighted pair.]{The 49 distinct hextile solutions, chiral pairs
grouped and the self-mirror marked, with the Hex9 pair highlighted. Of
49 = 24 pairs + 1 self-mirror, constraint A admits 18, constraint B
admits 8, and A ∩ B is the highlighted pair.\footnotemark{}}
\end{figure}
\footnotetext{Generated by
  \protect\texttt{experimental/halfhex\_further.py}; counts
  machine-verified by \protect\texttt{experimental/halfhex\_verify.py}.}

Axiom 6 resolves the remaining freedom: a single consistent chirality
choice, induced by the embedding into the reference ellipsoid frame,
selects one of the two. The orientation of the half-hexagon --- and with
it the entire hexagonal lattice across all 8 octants --- is determined:
no further choice remains.

This closes the enumeration. Each cell in the hierarchy is fixed by its
octant, its refinement level, and its index within that octant at that
level. No two cells share the same address; no address refers to more
than one cell. The hexagonal lattice is globally indexed with no
residual ambiguity. What follows from this is the subject of §9.

\begin{center}\rule{0.5\linewidth}{0.5pt}\end{center}

\subsection{9. Hex9 Cell Identity --- Structure as
Locator}\label{hex9-cell-identity-structure-as-locator}

The system defined by Axioms 1--9 and constructed through the preceding
steps is \textbf{Hex9}: a shifted-aperture-9 hexagonal grid on an
octahedral embedding of the reference ellipsoid, with a cell hierarchy
in which every cell has a distinct address, derived from simplicial
coherence requirements alone.

\begin{figure}
\centering
\includegraphics[width=0.92\linewidth,height=\textheight,keepaspectratio,alt={The seed solid and the 12 root cells. Left: the octahedron with each octant face creased into its three d\_cell facets (24 faces, the diploid form that names the d\_cell). Right: coloured per root x\_cell, hue by octahedral axis, light/dark for the mode-0/mode-1 halves --- at L0 every root cell is one of the 12 five-neighbour hexagons that meet at the octahedral vertices (appearing pentagonal once projected onto a reference ellipsoid). The faceting is illustrative; the cells live on the smooth ellipsoid.}]{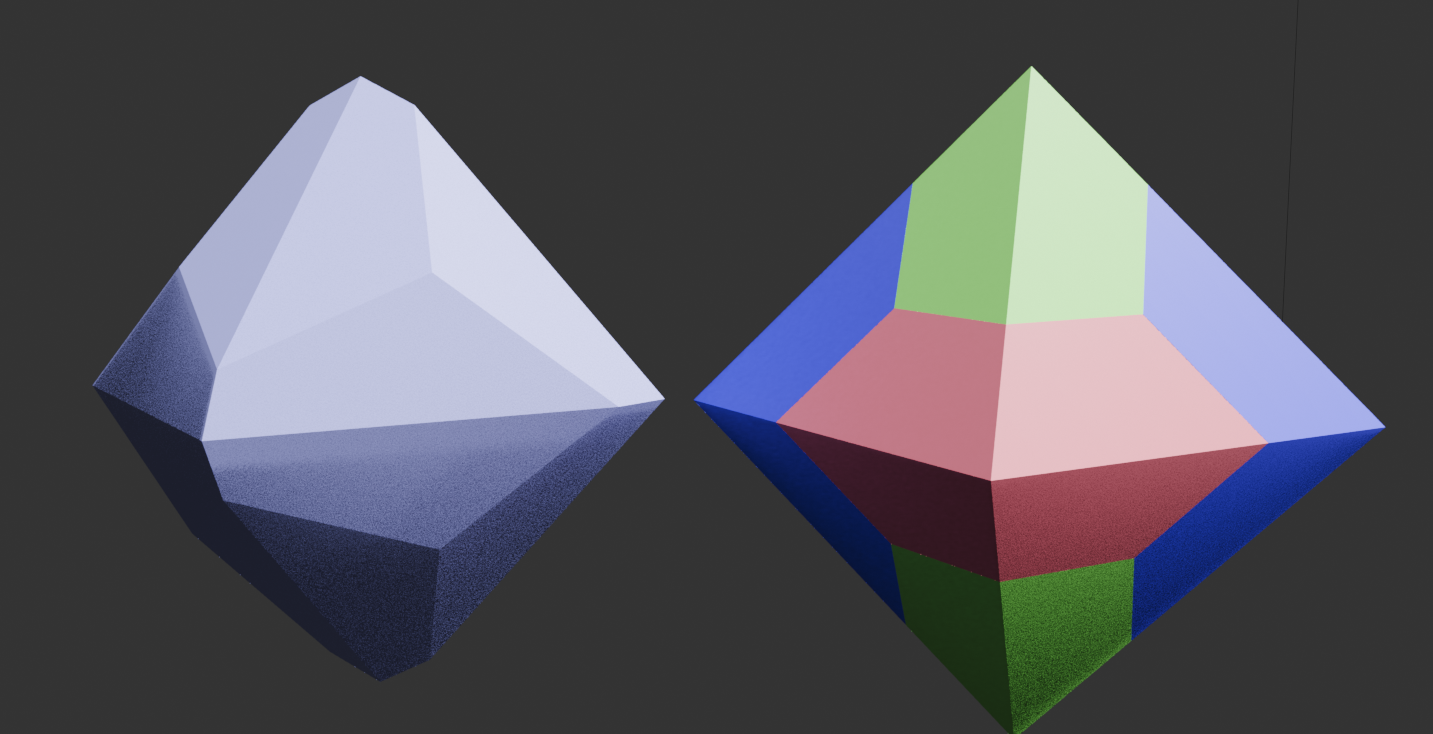}
\caption{The seed solid and the 12 root cells. Left: the octahedron with
each octant face creased into its three d\_cell facets (24 faces, the
diploid form that names the d\_cell). Right: coloured per root x\_cell,
hue by octahedral axis, light/dark for the mode-0/mode-1 halves --- at
L0 every root cell is one of the 12 five-neighbour hexagons that meet at
the octahedral vertices (appearing pentagonal once projected onto a
reference ellipsoid). The faceting is illustrative; the cells live on
the smooth ellipsoid.}
\end{figure}

Each Hex9 cell is identified by its octant (one of 8) and its path
through the refinement hierarchy. This pair is not a coordinate computed
from a prior reference system --- it is the cell's identity. The
bijection of Axiom 8 (Unique Cell Addressing) guarantees that this
identity encodes a specific, unambiguous geographic region. At each
finite refinement level the correspondence between addresses and cells
is exact: one address, one cell, no exceptions. Cell location is
recovered directly from combinatorial structure; no external coordinate
transformation is required to establish it.

Hex9 is a discrete spatial reference system in the sense of OGC Abstract
Specification Topic 21 (\citeproc{ref-ogc_topic21}{Open Geospatial
Consortium 2021}), which defines a DGGS as an integrated system
comprising a hierarchy of discrete global grids, spatio-temporal
referencing by zonal identifiers, and functions for quantization and
zonal query (its clause 4.13; Appendix B maps the vocabularies). What
distinguishes Hex9 within this class is the direction of derivation. The
conventional workflow --- select a coordinate reference system, then
design a grid over it --- is inverted. Here, the geometric coherence
requirements of §§1--8 define the structure; the structure constitutes
the locating system. No prior CRS is required as input.

This correspondence is discrete, not continuous. Hex9 cell identity
uniquely identifies a geographic region at each resolution. The
hierarchy of nested regions converges to a point as refinement
increases, but Hex9 does not provide continuous coordinates in the sense
of a projected CRS. For metric operations --- distances, areas,
interpolation --- a geometric realisation onto the reference ellipsoid
remains necessary, and is the subject of §11.

The enumeration completed in §8 is what makes self-contained location
possible. A grid with residual design freedom --- orientations,
apertures, or embeddings left as choices --- cannot serve as its own
locating system, because different choices produce different grids that
disagree on cell identity. Hex9 has no residual freedom. Every cell is
where it is because every constraint converges on that location.

\begin{center}\rule{0.5\linewidth}{0.5pt}\end{center}

\subsection{10. Addressing and
Continuity}\label{addressing-and-continuity}

\subsubsection{10.0 Notation --- the c/t/d/x grid
taxonomy}\label{notation-the-ctdx-grid-taxonomy}

The constructions of §§6--8 generate four overlapping grid spaces,
designated by single letters: \textbf{c}, \textbf{t}, \textbf{d},
\textbf{x}. The sections that follow use this vocabulary constantly;
this section fixes it. (The authoritative glossary, maintained alongside
the implementation, extends these definitions.)

\textbf{t\_cell} --- a triangular cell of the refined field: the working
unit of the simplicial carrier. Each t\_cell carries an intrinsic mode
(0 = ∇, 1 = Λ; §2) and subdivides into 9 child t\_cells at the next
level (§6). Within a parent context, a child t\_cell occupies one of 12
positional classes, written as a \textbf{region} id (0--11): 6 classes
are shared across both parent modes, 3 occur only under a mode-0 parent,
and 3 only under a mode-1 parent.

\textbf{c2} --- the label (0, 1, 2) of a t\_cell edge, assigned by edge
gradient: 0 = flat (horizontal), 1 = forward (positive slope), 2 = back
(negative slope). Adjacent triangles agree on the c2 value of their
common edge, and the c2 progression around any triangle is clockwise
regardless of mode. \emph{Etymology:} ``c2'' is shorthand for
\textbf{``colouring 2''} --- one of the two distinct three-colourings of
the wallpaper group \textbf{p31m} (Grünbaum \& Shephard 1987
(\citeproc{ref-grunbaum1987tilings}{Grünbaum and Shephard 1987}),
\emph{Tilings and Patterns}, §8.3 ``Color Pattern Types'', p. 433, type
IH38; PP25{[}3{]}₁/p31m{[}3{]}₁ and PP25{[}3{]}₂/p31m{[}3{]}₂ --- see
the \texttt{grunbaum\_shephard\_*} figures, the second with the Hex9
d\_cell overlay aligned). Note a \textbf{very early naming mismatch}:
despite the ``2'', the Hex9 c2 labelling actually corresponds to
Grünbaum \& Shephard's \emph{first} colouring, \textbf{p31m{[}3{]}₁},
not the second.

\textbf{c\_cell} --- a slot in the 96-position classifier grid: the raw
output of the three-family inequality classification of §10a. Twelve of
the 96 slots are the in-scope t\_cell classes; the remainder are out of
scope by construction.

\textbf{d\_cell} (half-hexagon) --- three t\_cells grouped by their
shared long edge. The c2 value of that long edge is the d\_cell's digit
(d\_dig ∈ \{0, 1, 2\}), and the d\_cell inherits the mode of its
t\_cells. The d\_cell is the fundamental domain of the tiling argument
in §8.

\textbf{x\_cell} (hexagon) --- one mode-0 d\_cell joined with one mode-1
d\_cell on their matching c2 edge. This is the public cell of the Hex9
grid: 12 root x\_cells cover the globe, and each x\_cell has 9 children.

\textbf{x\_dig} --- the digit (0--8) naming a child x\_cell within its
parent. Read it in ternary: the high trit encodes mode ownership (0 =
the child's mode-0 half is interior to the parent context; 1 = the
mode-1 half is interior; 2 = the child is \textbf{split}, straddling two
parents), and the low trit records the c2 orientation of the child's
long edge. The three split children per parent (x\_dig ∈ \{6, 7, 8\})
are the only cells with two valid parents (§10b).

\textbf{Lists and addresses} --- c\_list, t\_list, and d\_list are digit
sequences over strict single-parent trees: they compose left-to-right
and are prefix-sortable. An \textbf{x\_list} is the sequence of x\_digs;
because of the split cells it is resolved right-to-left. An
\textbf{x\_adr} is an x\_list plus a \textbf{tail} --- a single metadata
byte that resolves split-cell parentage and terminal state (§10b). The
tail is metadata only: it never participates in geometric computation.

\begin{figure}
\centering
\includegraphics[width=0.9\linewidth,height=\textheight,keepaspectratio,alt={Anatomy of one parent triangle: t\_cells, c2 edges, d\_cells, and the assembled x\_cells with x\_dig labels --- the c/t/d/x taxonomy in one picture.}]{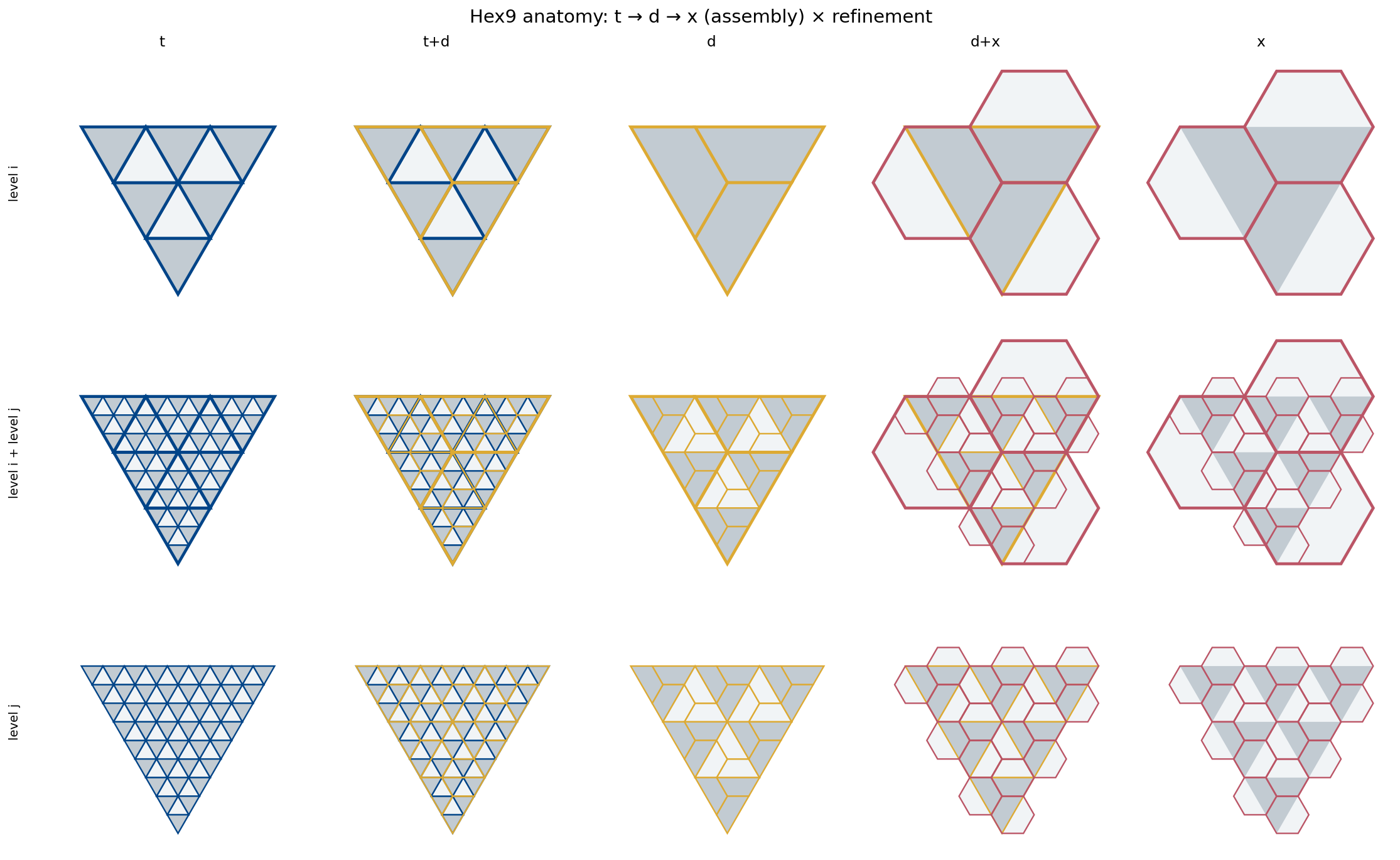}
\caption[Anatomy of one parent triangle: t\_cells, c2 edges, d\_cells,
and the assembled x\_cells with x\_dig labels --- the c/t/d/x taxonomy
in one picture.]{Anatomy of one parent triangle: t\_cells, c2 edges,
d\_cells, and the assembled x\_cells with x\_dig labels --- the c/t/d/x
taxonomy in one picture.\footnotemark{}}
\end{figure}
\footnotetext{Generated by \protect\texttt{examples/ex0400\_anatomy.py}.}

\subsubsection{10a. Identity as Locator}\label{a.-identity-as-locator}

A Hex9 address is a pair: an octant index (one of 8) and a refinement
path --- a sequence of x\_dig values recording which child x\_cell was
entered at each level of the hierarchy. An x\_adr of depth L identifies
exactly one x\_cell at level L on the reference ellipsoid.

The reversibility of this mapping rests on the construction. At each
level, a parent x\_cell contains exactly 9 child x\_cells, whose
arrangement is determined by the t\_cell → d\_cell → x\_cell sequence
established in §§6--8. Each x\_dig (0--8) selects one of those 9
children unambiguously. The refinement tree is a strict single-parent
structure for t\_cells and d\_cells; x\_cells inherit this property
except at the 3 split x\_cells per level, whose parentage is resolved by
the tail of the x\_adr.

\begin{figure}
\centering
\includegraphics[width=0.85\linewidth,height=\textheight,keepaspectratio,alt={The x-layer: a parent x\_cell and its 9 child x\_cells labelled by x\_dig (0--8), coloured by high-trit class (0--2 / 3--5 / 6--8). The three split children (6,7,8) straddle the parent boundary; there is no central child.}]{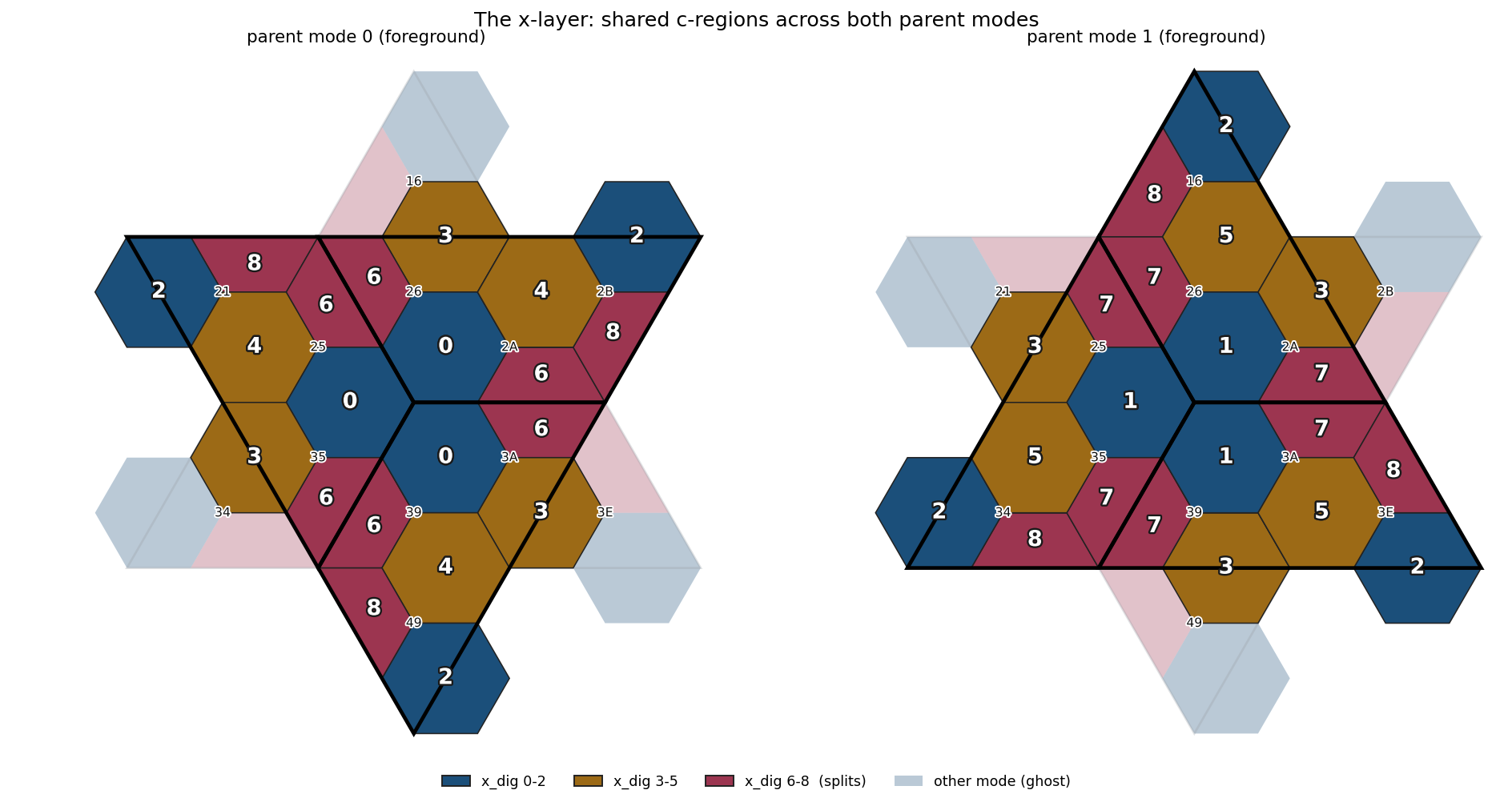}
\caption[The x-layer: a parent x\_cell and its 9 child x\_cells labelled
by x\_dig (0--8), coloured by high-trit class (0--2 / 3--5 / 6--8). The
three split children (6,7,8) straddle the parent boundary; there is no
central child.]{The x-layer: a parent x\_cell and its 9 child x\_cells
labelled by x\_dig (0--8), coloured by high-trit class (0--2 / 3--5 /
6--8). The three split children (6,7,8) straddle the parent boundary;
there is no central child.\footnotemark{}}
\end{figure}
\footnotetext{Generated by \protect\texttt{examples/ex0400\_anatomy.py}.}

Once the projection from the reference ellipsoid to the octant 2D plane
is complete, the encode direction (point → x\_adr) operates via a
sequence of linear inequality evaluations. Three families of parallel
lines partition the octant plane into the triangular grid:

\begin{itemize}
\tightlist
\item
  horizontal bands: \(y\) compared against fixed thresholds;
\item
  positive-slope bands: \(y - \sqrt{3}\,x\) compared against fixed
  thresholds;
\item
  negative-slope bands: \(y + \sqrt{3}\,x\) compared against fixed
  thresholds.
\end{itemize}

A point's t\_cell at a given level is determined by which band it
occupies on each family --- a small fixed number of comparisons, with no
geometric distance computation and no iterative search. At each
successive refinement level the thresholds scale by 1/3, maintaining
identical structure. The c\_cell (the 96-slot classifier combining
horizontal and slope bands) maps directly to a t\_cell, and from there
to the d\_cell and x\_cell via the construction of §§6--8.

\begin{figure}
\centering
\includegraphics[width=0.85\linewidth,height=\textheight,keepaspectratio,alt={The classifier (c-layer): three band families partition the octant plane; their indices compose the 96-slot c\_grid digit, which maps to a t\_cell and thence to d\_cell and x\_cell. Locating a point is band membership, not search.}]{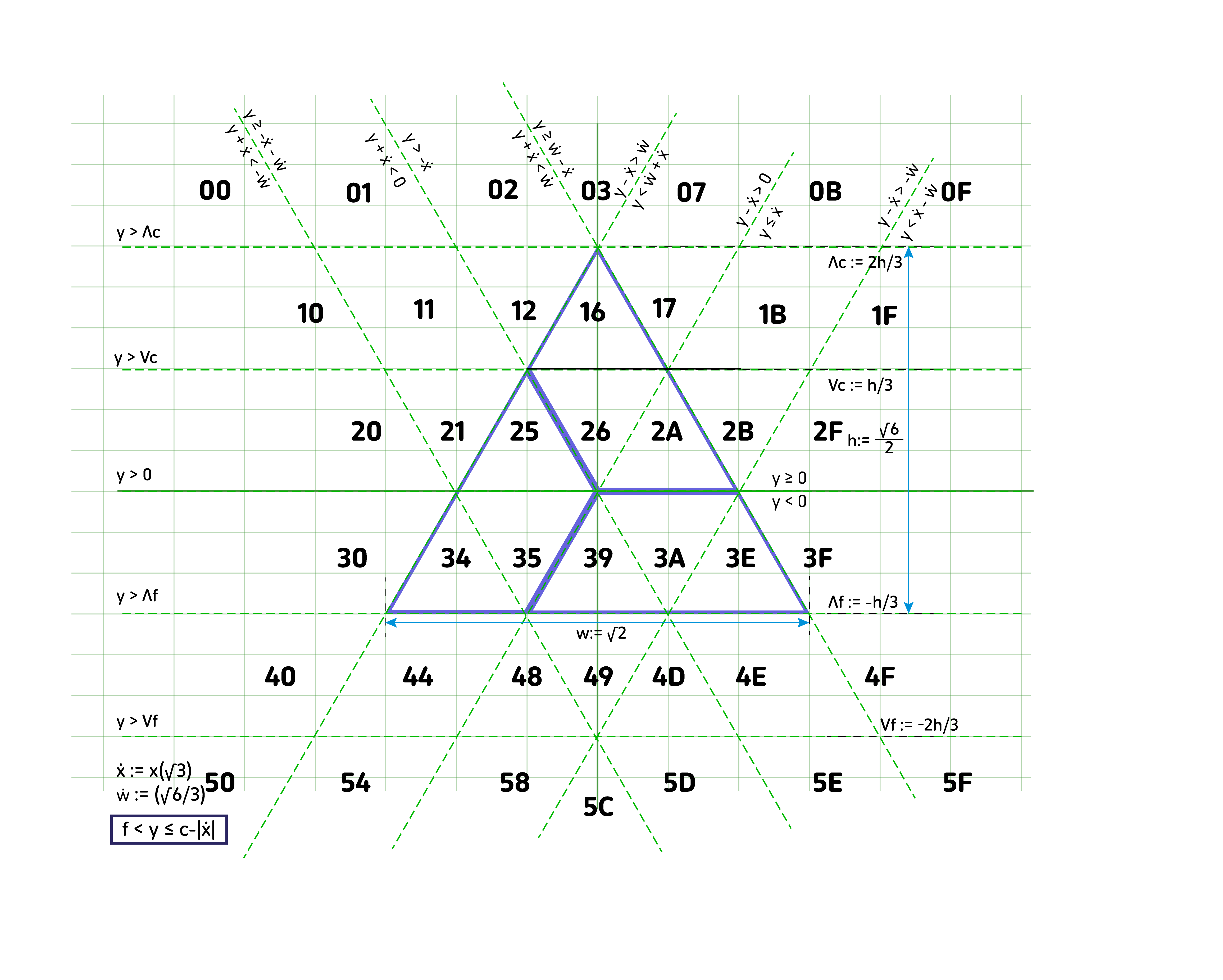}
\caption[The classifier (c-layer): three band families partition the
octant plane; their indices compose the 96-slot c\_grid digit, which
maps to a t\_cell and thence to d\_cell and x\_cell. Locating a point is
band membership, not search.]{The classifier (c-layer): three band
families partition the octant plane; their indices compose the 96-slot
c\_grid digit, which maps to a t\_cell and thence to d\_cell and
x\_cell. Locating a point is band membership, not
search.\footnotemark{}}
\end{figure}
\footnotetext{Generated from the \protect\texttt{addressing.py}
  classifier.}

The mapping operates in both directions. Given an x\_adr, the
corresponding geographic region is recovered by tracing the digit
sequence from the octant root: each x\_dig selects a child x\_cell whose
boundary is determined by the refinement geometry. Given any point on
the ellipsoid, its x\_adr at level L is recovered by projecting to the
octant plane and evaluating the inequality sequence down to depth L.

Both directions are exact at each finite level. Encode produces no
approximation: every point belongs to exactly one x\_cell at every
level. Decode recovers a cell whose geographic extent is precisely
determined by the octahedral embedding and the refinement geometry.

This reversibility qualifies Hex9 as a locating system rather than
merely an indexing scheme. Location is not inferred from the address ---
it is recovered from it by retracing the construction.

\subsubsection{10b. Identity as Key}\label{b.-identity-as-key}

A Hex9 address is also a spatial key --- a string over a small digit
alphabet that can be stored, compared, and sorted without reference to
any geometric structure. This makes Hex9 addresses directly usable as
database keys, hash keys, or binning primitives.

The key property is that prefix order corresponds to containment. All
cells whose address begins with a given prefix σ are contained within
the cell identified by σ. This means spatial containment queries reduce
to prefix comparisons: to find all level-L cells within a given level-K
region (L \textgreater{} K), select all addresses sharing that region's
prefix. No geometric computation is needed.

Spatial joins between two datasets reduce similarly: two observations
share a cell at level K if and only if their addresses share a common
prefix of length K. Aggregation across levels is achieved by truncating
addresses to the desired depth. These operations are efficient and
require no coordinate arithmetic.

Because the address space is defined by the refinement structure rather
than by a numerical coordinate grid, there are no edge effects, no
wrap-around anomalies, and no cells that straddle index boundaries.
Every cell has exactly one address; every address identifies exactly one
cell. The key space is clean.

This last claim is grounded in the address tail. The x\_list alone is
not always sufficient: at each level, 3 of the 9 child x\_cells (those
with x\_dig in \{6,7,8\}) straddle the c2 boundary between two adjacent
d\_cells and have two valid parents. Beyond this, two cells with
distinct terminal regions can produce identical digit sequences when
those regions generate the same hex digit from different parent c2
contexts.

The key tail carries two fields that resolve these ambiguities. The
first is p\_c2 --- the parent c2 of the terminal region. A concrete
instance: hex digit 6 at the terminal level arises from both region 9
(mode 1, parent c2 = 0) and region 6 (mode 0, parent c2 = 2) within the
same parent context. Both paths produce an identical digit body; without
p\_c2 they collide as keys. With p\_c2 = 0 or p\_c2 = 2 respectively,
the two addresses are distinct. The second field is r\_mo --- the root
octant's net\_mode. Two octants of opposite mode can produce the same
root hex digit; without r\_mo the decoder cannot recover which octant
the address originates from, and cells in distinct geographic regions
would share the same key.

The full reversible tail additionally carries p\_mo and h. The p\_mo
field records the actual parent mode of the terminal region; for split
x\_cells, this may differ from the key tail's canonical mode-0
assumption. Without p\_mo, the decoder recovers the mode-0 parent's
representative, not the exact terminal cell. The h field identifies the
terminal d\_cell (one of 12, bits 3--0); without it, reconstruction
returns the x\_cell centre. With h, the precise d\_cell centroid is
recovered. The reversible tail is fully invertible: the address body
together with the reversible tail uniquely and exactly determines both
the geographic region and a representative point within it.

The tail fields in summary:

{\def\LTcaptype{none} % do not increment counter
\begin{longtable}[]{@{}
  >{\raggedright\arraybackslash}p{(\linewidth - 6\tabcolsep) * \real{0.2500}}
  >{\raggedright\arraybackslash}p{(\linewidth - 6\tabcolsep) * \real{0.2500}}
  >{\raggedright\arraybackslash}p{(\linewidth - 6\tabcolsep) * \real{0.2500}}
  >{\raggedright\arraybackslash}p{(\linewidth - 6\tabcolsep) * \real{0.2500}}@{}}
\toprule\noalign{}
\begin{minipage}[b]{\linewidth}\raggedright
Field
\end{minipage} & \begin{minipage}[b]{\linewidth}\raggedright
Bits
\end{minipage} & \begin{minipage}[b]{\linewidth}\raggedright
Carries
\end{minipage} & \begin{minipage}[b]{\linewidth}\raggedright
Without it
\end{minipage} \\
\midrule\noalign{}
\endhead
\bottomrule\noalign{}
\endlastfoot
p\_c2 & 2 & parent c2 of the terminal region & distinct cells collide as
keys (same digit body from different parent c2 contexts) \\
r\_mo & 1 & root octant net mode & octant unrecoverable from the root
digit; cross-octant key collisions \\
p\_mo & 1 & actual parent mode of the terminal region (reversible tail
only) & decoder returns the canonical mode-0 representative, not the
exact cell \\
h & 4 & terminal d\_cell id, 0--11 (reversible tail only) &
reconstruction returns the x\_cell centre, not the exact d\_cell
centroid \\
\end{longtable}
}

Both tails are invertible: the key (bin) tail round-trips to the bin
cell's representative point, the full tail (all four fields) to the
terminal cell's. The difference is \emph{which} cell's representative is
recovered --- never the original source point --- not whether one is
recoverable at all.

One caution follows directly. Truncating an address to length K always
identifies a valid ancestor at the corresponding level, but not
necessarily the \emph{canonical} one: a split cell encoded under its
mode-1 parent truncates into the mode-1 lineage. Binning by naive
prefix-cutting therefore silently produces two bins for the same cell
whenever non-canonical addresses are present. The correct operation
derives the canonical ancestor via the tail before truncating:
prefix-cutting is exact for resolution identification; canonical
ancestry requires the tail.

A worked example: central London. The Prime Meridian is a c2 boundary in
this region. At level 4, the cells around Greenwich sit in three
adjacent hexagons with non-adjacent prefixes:

{\def\LTcaptype{none} % do not increment counter
\begin{longtable}[]{@{}ll@{}}
\toprule\noalign{}
L4 digit body (lineage) & Location \\
\midrule\noalign{}
\endhead
\bottomrule\noalign{}
\endlastfoot
43483 & corner of north London \\
43486 & east of Greenwich \\
43527 & west of Greenwich \\
\end{longtable}
}

43527 appears to have jumped lineage: it belongs to 4352 (south-west
England) despite being geographically adjacent to 43486 in 4348 (east
England). The jump is not an anomaly --- it is the visible signature of
the split digits. Digits 6--8 carry high ternary trit 2: thesecells
straddle the d\_cell boundary, and the canonical mode-0 parent
convention places geographically adjacent cells on opposite sides of
that boundary into different canonical lineages, exactly as the
construction requires.

\begin{figure}
\centering
\includegraphics[width=0.9\linewidth,height=\textheight,keepaspectratio,alt={The split-cell lineage jump over central London and the Thames estuary (Equal Earth, EPSG:8857). Hexagons are annotated by digit body --- the lineage path, §10c; the bold 4348 and 4352 mark the two level-3 parent cells, and the heavy line is their boundary. Cell 43527 lies west of Greenwich yet belongs to the 4352 lineage, while its geographic neighbours 43486 and 43483 belong to 4348 --- the canonical mode-0 convention placing adjacent cells across the c2 (Prime-Meridian) boundary into different lineages, exactly as §10b describes.}]{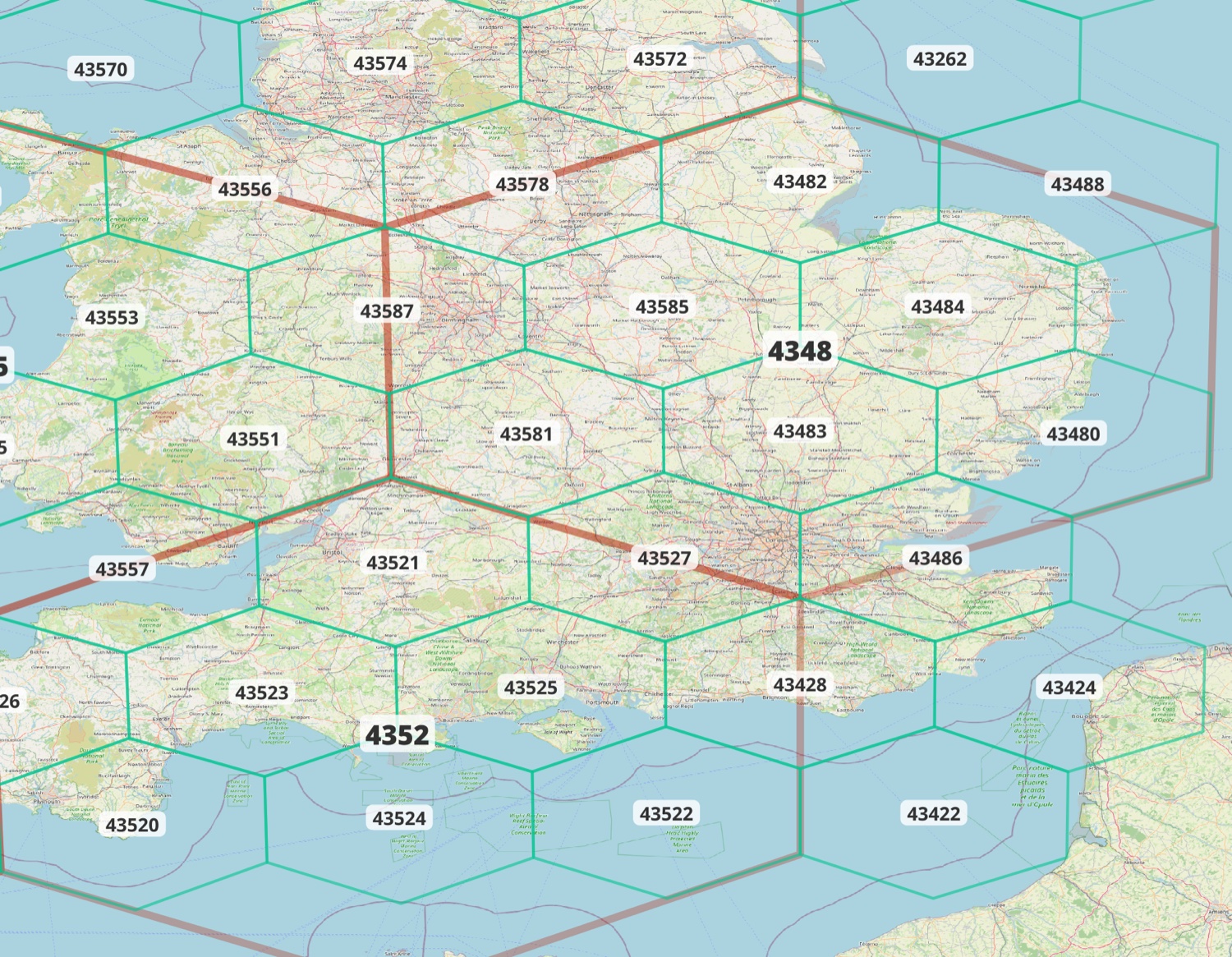}
\caption[The split-cell lineage jump over central London and the Thames
estuary (Equal Earth, EPSG:8857). Hexagons are annotated by digit body
--- the lineage path, §10c; the bold \textbf{4348} and \textbf{4352}
mark the two level-3 parent cells, and the heavy line is their boundary.
Cell \textbf{43527} lies west of Greenwich yet belongs to the
\textbf{4352} lineage, while its geographic neighbours \textbf{43486}
and \textbf{43483} belong to \textbf{4348} --- the canonical mode-0
convention placing adjacent cells across the c2 (Prime-Meridian)
boundary into different lineages, exactly as §10b describes.]{The
split-cell lineage jump over central London and the Thames estuary
(Equal Earth, EPSG:8857). Hexagons are annotated by digit body --- the
lineage path, §10c; the bold \textbf{4348} and \textbf{4352} mark the
two level-3 parent cells, and the heavy line is their boundary. Cell
\textbf{43527} lies west of Greenwich yet belongs to the \textbf{4352}
lineage, while its geographic neighbours \textbf{43486} and
\textbf{43483} belong to \textbf{4348} --- the canonical mode-0
convention placing adjacent cells across the c2 (Prime-Meridian)
boundary into different lineages, exactly as §10b
describes.\footnotemark{}}
\end{figure}
\footnotetext{Rendered in QGIS over an OpenStreetMap backdrop; hex
  boundaries and addresses from \protect\texttt{hhg9}.}

\subsubsection{10c. Identity as Label}\label{c.-identity-as-label}

A label is the human-readable serialisation of the key: the digit body
followed by the key-tail nibble, written
\texttt{\textless{}body\textgreater{}.\textless{}tail\textgreater{}} ---
\texttt{32343.2} names the level-4 cell whose canonical digit path is
32343 under key tail 2. The form is implemented by \texttt{h9\_label}
and its inverse \texttt{h9\_from\_label}; labels and keys are in
bijection, so a label carries exactly the identity of §10b, in a form
that can be printed, sorted, and compared by eye.

The digit body alone --- the part before the dot --- is a \emph{lineage
path}, not an identity. It is readable and prefix-meaningful: truncating
it walks up a lineage, and two bodies sharing a prefix share the
corresponding ancestor context. But it fails as a name in both
directions (§10b): a split cell owns two valid bodies, one per parent
lineage, and two distinct cells can produce the same body from different
parent contexts --- the p\_c2 and r\_mo collisions. The tail nibble is
what closes both gaps. A string without it should be read as a path
through the hierarchy, never as a cell.

What makes labels uniform is the canonical fold of §10b, not a
suppression of structure: after canonicalisation every cell is presented
through a mode-0 terminal parent, so a label needs no mode flag, lookup
table, or supplementary field to interpret. The bipartite mode structure
remains present in the refinement geometry and is recoverable from the
key. A label therefore identifies a specific geographic region at a
specific resolution, in one canonical spelling --- nothing more is
needed, and nothing is omitted.

\subsubsection{10d. Adjacency from Refinement
Paths}\label{d.-adjacency-from-refinement-paths}

Two x\_cells are adjacent if they share a boundary edge on the reference
ellipsoid. In Hex9, adjacency is recoverable from the refinement
structure because the x\_cell geometry is fully determined by the
t\_cell → d\_cell → x\_cell construction.

Within a parent x\_cell, the 9 child x\_cells tile the parent's region.
Their shared edges are the c2 edges of the underlying d\_cells --- the
same long-edge alignment that the orientation selection of §8 fixed
globally. Adjacency is realised as a fixed, finite lookup: every cell
has exactly three neighbours, one per c2 value, given by a constant
table keyed by (cell, parent mode, c2). The digit structure mirrors
this: each split digit k+6 (k ∈ \{0, 1, 2\}) names the pair of d\_cells
flanking interior child k, and the low ternary trit of every x\_dig
records the c2 orientation of the cell's long edge.

\begin{figure}
\centering
\includegraphics[width=0.85\linewidth,height=\textheight,keepaspectratio,alt={Sibling adjacency within a parent: each child t\_cell edge is internal (blue, shared with a sibling) or external (red, cross-parent); a child's adjacency class is its red-edge count (interior 0, mid-edge 1, vertex 2).}]{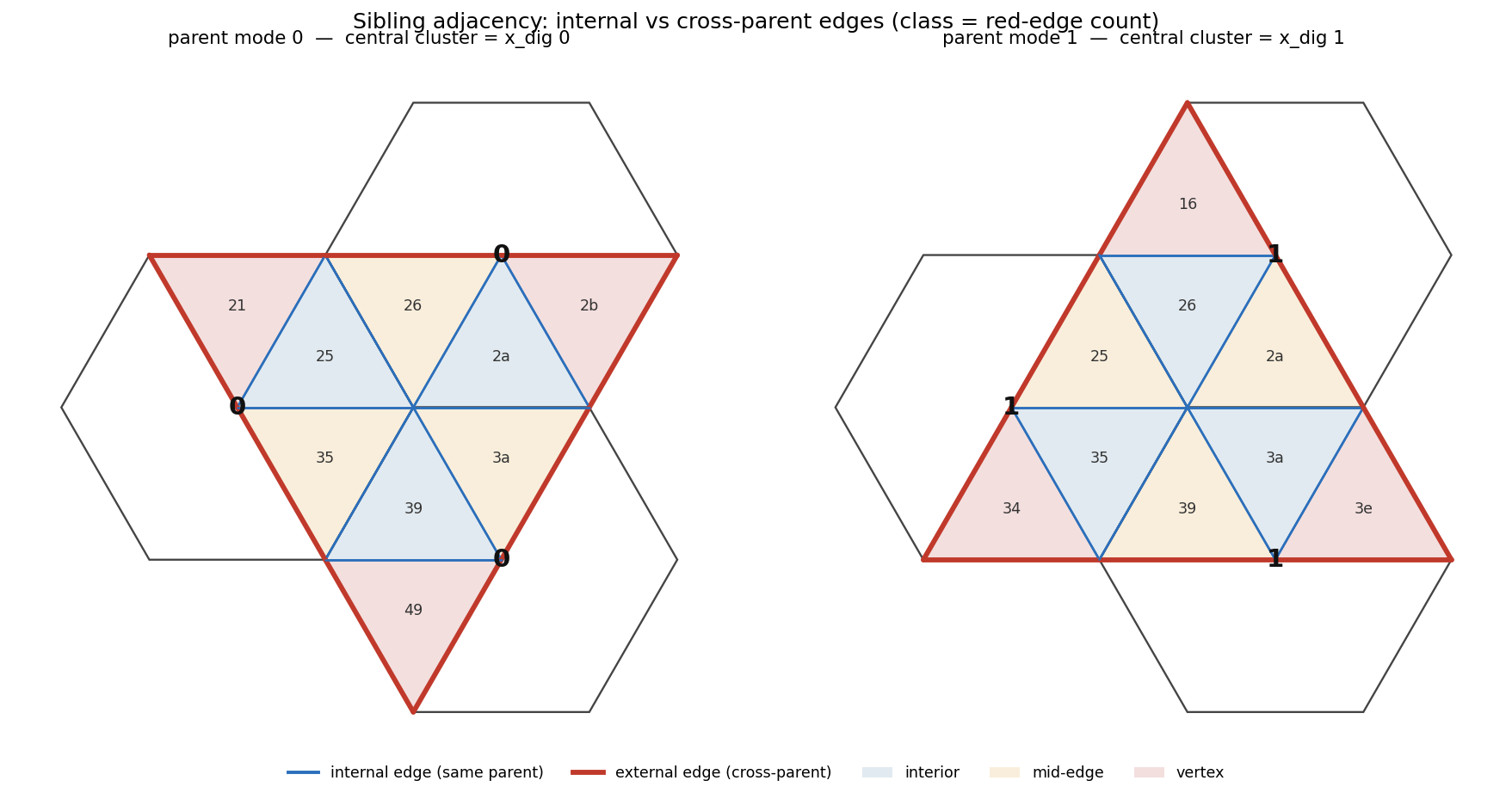}
\caption[Sibling adjacency within a parent: each child t\_cell edge is
internal (blue, shared with a sibling) or external (red, cross-parent);
a child's adjacency class is its red-edge count (interior 0, mid-edge 1,
vertex 2).]{Sibling adjacency within a parent: each child t\_cell edge
is internal (blue, shared with a sibling) or external (red,
cross-parent); a child's adjacency class is its red-edge count (interior
0, mid-edge 1, vertex 2).\footnotemark{}}
\end{figure}
\footnotetext{Generated by \protect\texttt{examples/ex0400\_anatomy.py};
  classes verified against \protect\texttt{region.py}'s neighbour
  builder.}

The same table classifies every cell's relationship to its parent
boundary. Within each parent, cells fall into three classes: interior
cells, whose neighbours all share the parent; mid-edge cells, with
exactly one neighbour in an adjacent parent; and vertex cells, with two.
For the latter classes the lookup flags the crossing and identifies the
relevant child in the neighbouring parent --- a computable operation on
the address structure, stepping up one level in the refinement tree and
applying the same c2 edge relationships that govern interior adjacency.
Where the neighbouring parent lies across an octant seam, the octant
congruence of §5 reduces the hop to an octant-index lookup composed with
the y-reflection.

Adjacency across octant seams follows from the d\_cell c2 alignment at
octant boundaries. The orientation selection of §8 ensures that d\_cells
at an octant edge meet long-edge to long-edge with the d\_cells of the
neighbouring octant. The x\_cells that form across this boundary are
assembled by the same d\_cell join rule as everywhere else.
Seam-crossing adjacency is therefore not a special case --- it is the
same c2-edge query applied at the octant boundary.

In each case adjacency is a finite computation on the combinatorial
structure of the refinement tree and the c2 edge table. No geometric
distance query is needed, and no location in the grid requires a
different procedure from any other.

\subsubsection{10e. Continuity}\label{e.-continuity}

At any finite level L, a Hex9 address identifies a cell of finite
geographic extent. As L increases, cell diameter decreases by a factor
of 3 at each level, converging toward zero. An infinite address sequence
--- a path carried to all levels of the hierarchy --- therefore defines
a nested sequence of cells whose diameters tend to zero.

The reference ellipsoid is a compact metric space. By Cantor's
intersection theorem, a nested sequence of closed regions with diameters
converging to zero has exactly one point in its intersection. An
infinite Hex9 address sequence identifies exactly that point: not a
region, but a location on the ellipsoid.

This convergence lets a function recover position from an address to
arbitrary precision: decoding an address of growing length yields a
sequence of points converging to a unique location on the ellipsoid
(Appendix A). A finite address identifies a region; a sufficiently long
address identifies a location to any required tolerance.

This is not the same as being a coordinate reference system in the
strict sense of ISO 19111, and we do not claim it is. Two limitations
are intrinsic. First, the address alphabet is discrete: addresses form a
totally disconnected sequence space, not a continuum, so they are not
coordinates in the real-valued sense. Second, the point → address map is
discontinuous on the measure-zero set of d\_cell seams (the split-cell
boundaries of §10b): arbitrarily close points on opposite sides of a
seam receive addresses that differ in their leading digits. ISO 19111
presupposes continuous coordinates, and Hex9 does not meet that
requirement.

What Hex9 offers is better described, by analogy with its quasi-authalic
geometry, as \textbf{quasi-continuous}: position is recoverable from the
address by a function, to arbitrary precision, everywhere except on a
measure-zero seam set, and the cell hierarchy converges to points rather
than terminating at a finite floor. We are not aware of another DGGS
whose zonal identifier doubles as a position-recovery coordinate in this
way, though we do not claim the property is unique. At any fixed finite
resolution Hex9 remains a discrete system --- cells are regions, the
bijection is between addresses and regions --- and the quasi-continuous
behaviour emerges only in the limit. This is consistent with §9: Hex9 is
a discrete spatial reference system that approaches, but does not
attain, a continuous one.

\subsubsection{10f. Seams and Valence
Defects}\label{f.-seams-and-valence-defects}

The Euler characteristic of S² requires that any triangulation of the
globe carry topological defects. In Hex9 these are absorbed at the six
octahedral vertices --- each a 4-valent point of the triangulation,
where the hexagonal grid carries its defect as the five-neighbour cells
of §12. Understanding why neither these defects nor octant seams require
special handling requires tracing the constructive sequence: t\_cells →
d\_cells → x\_cells.

At each refinement level, each triangular face (t\_cell) is subdivided
into 9 child t\_cells. Three adjacent t\_cells --- grouped by their
shared long edge (c2 edge) --- form a half-hexagon (d\_cell). A d\_cell
carries an intrinsic mode (0 or 1). One mode-0 d\_cell and one mode-1
d\_cell, joined on their matching c2 edge, form a hexagonal cell
(x\_cell). The hexagonal grid emerges entirely from this sequence; no
independent hex tiling is assumed.

\emph{(The t\_cell → d\_cell → x\_cell constructive sequence is the
anatomy figure of §10.0.)}

At an octant seam, the d\_cell c2 alignment ensures that d\_cells on
either side of the shared octant edge meet long-edge to long-edge. This
alignment is not enforced separately --- it is the consequence of the
orientation selection of §8, which chose precisely the arrangement in
which c2 edges align at every boundary. The x\_cells that straddle the
seam are formed by the same d\_cell joining rule that applies everywhere
else. The seam is a boundary in the refinement tree, not a discontinuity
in the construction.

At an octahedral vertex, four octant faces meet --- the six seed
vertices are 4-valent, where every interior vertex of the refined
triangulation is 6-valent. The surrounding d\_cells at this vertex have
c2=0 --- the flat (horizontal) edge --- as their shared long edge. This
is the c2=0 convention: not a patch, but the consequence of the same
vertex-loop transport closure established in §4. Even valence at the
octahedral vertices (valence 4) satisfies the mode transport condition;
the surrounding x\_cells are formed by the same d\_cell join rule. The
cells meeting at the vertex are the five-neighbour hexagons of §12,
structurally identical in construction to every other x\_cell --- each a
hexagon that, projected onto the ellipsoid, presents one collinear
corner and so appears pentagonal.

In both cases the construction proceeds without branching. The t\_cell →
d\_cell → x\_cell sequence applies uniformly across the entire globe ---
at seams, at defect vertices, and in the interior. The absence of
special cases follows from the coherence of the construction, not from
exception handling added afterward.

\begin{center}\rule{0.5\linewidth}{0.5pt}\end{center}

\subsection{11. Geometric Realisation (AK +
Warp)}\label{geometric-realisation-ak-warp}

Sections 1--10 establish the Hex9 grid as a combinatorial object: mode
transport, t\_cells, d\_cells, x\_cells, the refinement hierarchy, and
the address structure are all defined without reference to any specific
map projection or reference body. The geometric realisation is a
separable step that places this abstract structure onto the WGS84
reference ellipsoid.

Three concerns are independent: (1) the combinatorial structure and
hierarchy (§§1--10); (2) the base projection from octahedron to
ellipsoid; (3) the area correction. Each can be understood, improved, or
substituted without disturbing the others. The warp is
ellipsoid-specific; the grid is not.

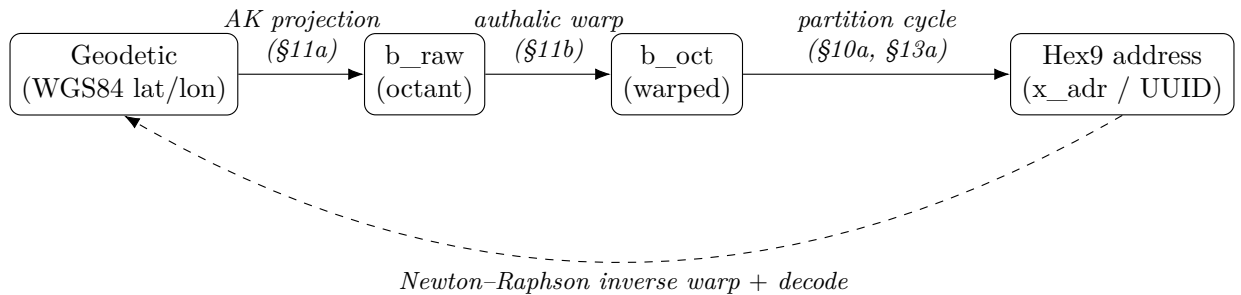
\begin{figure}[h]
\centering
\resizebox{\textwidth}{!}{%
\begin{tikzpicture}[node distance=8mm and 16mm, >={Latex[length=2mm]},
  box/.style={draw, rounded corners, align=center, font=\small,
    minimum height=10mm, inner sep=4pt},
  op/.style={font=\footnotesize\itshape, align=center}]
  \node[box] (g)   {Geodetic\\(WGS84 lat/lon)};
  \node[box, right=of g]   (raw) {b\_raw\\(octant)};
  \node[box, right=of raw] (oct) {b\_oct\\(warped)};
  \node[box, right=34mm of oct] (adr) {Hex9 address\\(x\_adr / UUID)};
  \draw[->] (g)   -- node[op, above]{AK projection\\(\S11a)} (raw);
  \draw[->] (raw) -- node[op, above]{authalic warp\\(\S11b)} (oct);
  \draw[->] (oct) -- node[op, above]{partition cycle\\(\S10a, \S13a)} (adr);
  \draw[->, dashed] (adr.south) to[out=-150,in=-30]
    node[op, below]{Newton--Raphson inverse warp $+$ decode} (g.south);
\end{tikzpicture}}
\caption{The geometric-realisation pipeline (\S11). A WGS84 geodetic coordinate
reaches a combinatorial Hex9 address through the AK base projection, the authalic
warp, and the partition cycle; the inverse (dashed) retraces the chain with a
Newton--Raphson warp inversion. The three concerns --- combinatorial hierarchy,
base projection, area correction --- are independent and separately substitutable.}
\end{figure}

\subsubsection{11a. The Base Projection}\label{a.-the-base-projection}

The AK octahedral projection maps each of the 8 octant faces to the
corresponding region of the reference ellipsoid. Its forward formula,
designed analytically by Anders Kaseorg
(\citeproc{ref-kaseorg_octahedral}{Kaseorg 2025}) from a force-directed
dataset, applies a tangent substitution to each octant coordinate and
couples the three axes with a fourth-root term modulated by a parameter
α ≈ 3.2278. This coupling partially compensates for the area asymmetry
between mode-0 and mode-1 triangular cells that arises from the
non-uniform Jacobian of the octant face.

The projection is smooth and has an analytical Jacobian --- properties
that make it suitable as the inner layer beneath the Sinkhorn warp. The
8 octahedral vertices coincide exactly with ellipsoidal surface points;
no projection computation is needed at those locations. The 8 octant
faces are mutually equivalent under a y-coordinate reflection, so one
projection function serves all octants.

No closed-form inverse exists. The backward pass --- ellipsoid to octant
--- is realised by numerical root-finding; the fast path is a guarded
Gauss--Newton iteration on the analytic Jacobian (§13f), with an exact
beam search as the reference and as the fallback near seams and
vertices. The forward map is smooth and injective on each octant, so an
inverse is guaranteed to exist; the numerical method is an
implementation choice, not a structural requirement.

Without further correction, the AK projection introduces roughly ±20\%
area deviation across the octant face --- larger apparent cells near
octant corners, smaller near the centre.

A second, subtler artefact is an inter-mode area bias. The octant face
is a right isosceles triangle, so its three corners are not
geometrically equivalent and the AK Jacobian is not constant over the
face; mode-0 and mode-1 triangles sample that non-uniform field at
systematically different centroid positions. Measured by geodesic area
on WGS84, the mode means differ by 0.394\% at L4 and 0.131\% at L5,
shrinking roughly threefold per refinement level (projected ≈ 0.04\% at
L6). The coupling parameter α is not the cause --- removing it (α = 0)
worsens both the global deviation (σ ≈ 7.6\% → ≈ 17\%) and the asymmetry
--- and the Sinkhorn warp, operating at the hexagon level, cannot
rebalance areas between the two halves of a hexagon, so the residual
carries through to the warped result at slightly reduced magnitude. At
L5 and finer it is negligible for practical use.

\subsubsection{11b. The Authalic Warp}\label{b.-the-authalic-warp}

The warp corrects the area deviation left by the base projection.
(``Authalic'' means equal-area: the warp is not itself a projection but
the corrective deformation that renders the composite octahedral map
area-true.) It is derived by Sinkhorn optimal transport
(\citeproc{ref-cuturi2013sinkhorn}{Cuturi 2013}): treating the L4 (or
L5) triangle vertices as a discrete mass distribution on the octant, the
Sinkhorn iteration finds the minimal-displacement redistribution that
equalises projected cell areas against the geodesic areas they subtend
on WGS84. The result is a displacement field --- corrections in b\_oct
coordinates --- rather than absolute positions, which improves numerical
conditioning.

The displacement field is precomputed once per ellipsoid and stored. At
runtime, it is applied via a Clough-Tocher interpolant
(\citeproc{ref-clough1965finite}{Clough and Tocher 1965}) (C1
continuous). The inverse warp uses the forward interpolant to obtain an
initial estimate, then refines by Newton-Raphson to a tolerance of 10⁻¹⁴
in b\_oct (barycentric) units. The geodetic round-trip g → b\_oct → g,
measured on WGS84 at validation points that include a near-pole location
(89.99°N) and the Greenwich seam, returns to within 1.8 nm of the
original position --- many orders of magnitude below any geodetic
relevance, and comfortably under the 7 nm design threshold.

The achieved area uniformity (L5, all 708,588 hexagons, WGS84,
production warp file, geodesic areas): mean deviation exactly 0.000\%
(confirming closure: the cell areas sum to the ellipsoid surface area);
area deviation min −3.57\%, max +4.80\%; mean absolute deviation
0.001\%; log-ratio standard deviation 1.99×10⁻⁴. Half of all cells are
within 0.0002\% of ideal area; 99\% within 0.0044\%; 99.99\% within
0.43\%. The extreme values are a balanced ±4\% pair affecting a very
small number of cells immediately adjacent to the six octahedral
vertices; the bulk distribution is highly uniform. The warp is
quasi-authalic rather than strictly authalic. A strictly authalic
projection constrains only the Jacobian determinant, permitting severe
shear and cell elongation. The optimal-transport derivation implicitly
regularises against shear by minimising displacement: the result trades
a small area residual for a smooth, low-distortion displacement field.

\begin{figure}
\centering
\includegraphics[width=0.75\linewidth,height=\textheight,keepaspectratio,alt={Per-hex area deviation around the north-pole vertex, colour scale capped at ±0.25\%. The tight cap reveals the residual structure: a quiet near-white interior (the equal-area result), the vertex bloom whose innermost cells reach the ±4--5\% extremes and saturate the scale, and the gentle grading that runs out along the two visible octant seams. Every cell is one L5 hexagon.}]{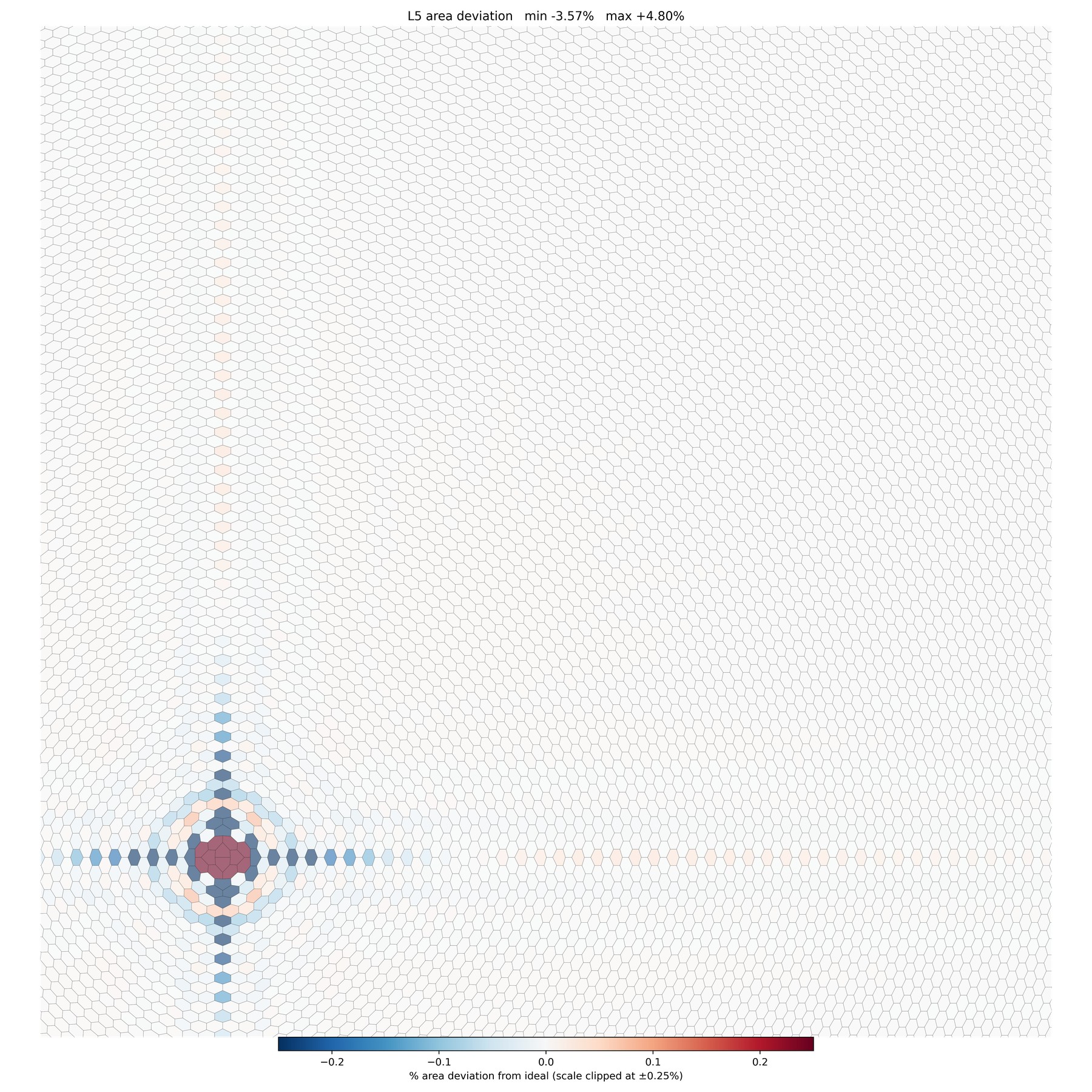}
\caption[Per-hex area deviation around the north-pole vertex, colour
scale capped at ±0.25\%. The tight cap reveals the residual structure: a
quiet near-white interior (the equal-area result), the vertex bloom
whose innermost cells reach the ±4--5\% extremes and saturate the scale,
and the gentle grading that runs out along the two visible octant seams.
Every cell is one L5 hexagon.]{Per-hex area deviation around the
north-pole vertex, colour scale capped at ±0.25\%. The tight cap reveals
the residual structure: a quiet near-white interior (the equal-area
result), the vertex bloom whose innermost cells reach the ±4--5\%
extremes and saturate the scale, and the gentle grading that runs out
along the two visible octant seams. Every cell is one L5
hexagon.\footnotemark{}}
\end{figure}
\footnotetext{Generated by
  \protect\texttt{examples/ex0081w\_warped\_authalics.py}
  (\protect\texttt{snow\_globe}, \protect\texttt{lim\_pct=0.25}).}

\begin{figure}
\centering
\includegraphics[width=0.92\linewidth,height=\textheight,keepaspectratio,alt={Per-hex area deviation, Mollweide, on a scale clipped at −0.31\%/+0.26\% --- the p1--p99 band (pattern view). The clip makes the spatial structure legible: a quiet near-white interior with the octant seam skeleton and six vertex blooms picked out. The colourbar is clipped and does not reach the true extremes (min −3.57\%, max +4.80\%).}]{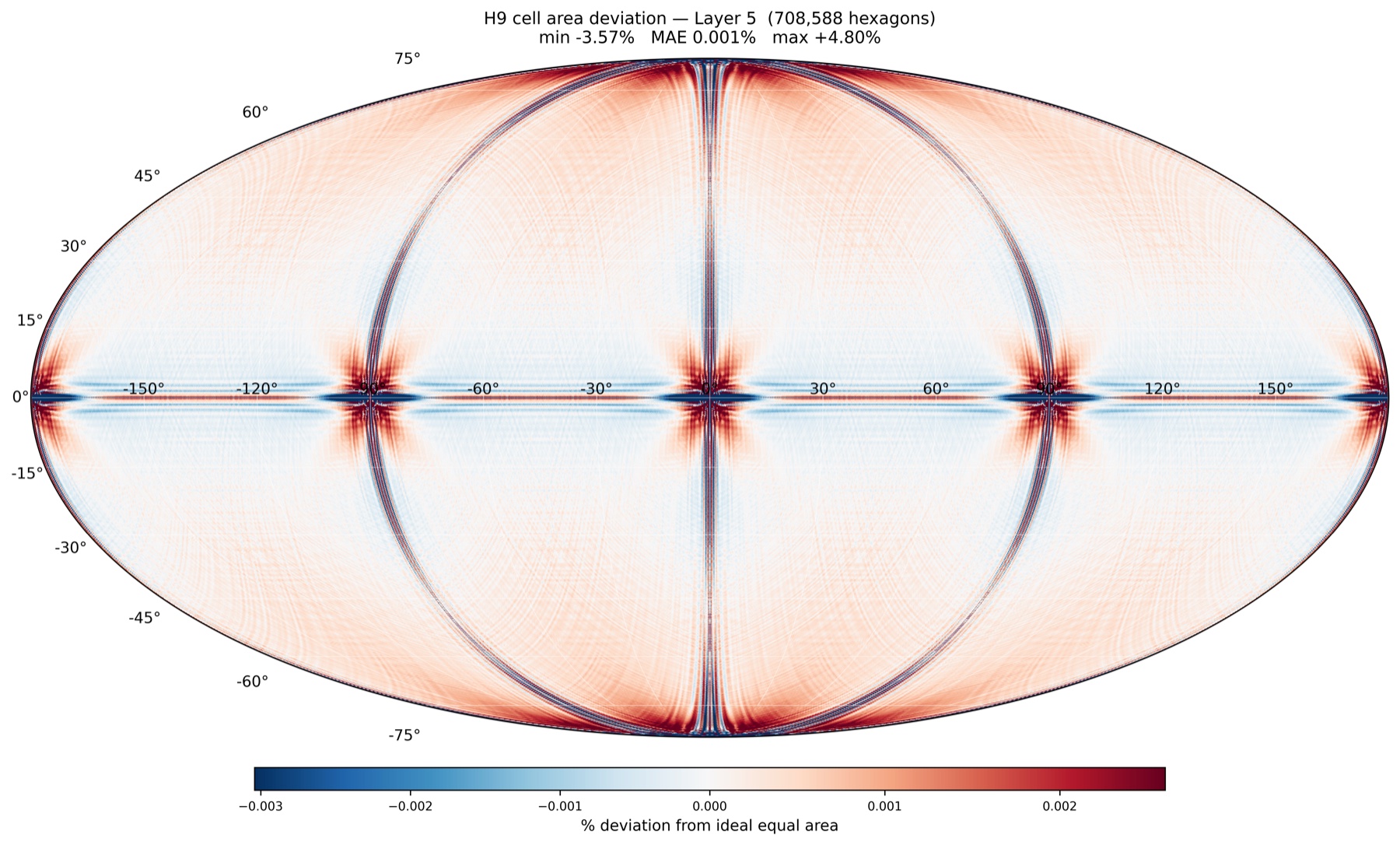}
\caption[Per-hex area deviation, Mollweide, on a scale clipped at
−0.31\%/+0.26\% --- the p1--p99 band (pattern view). The clip makes the
spatial structure legible: a quiet near-white interior with the octant
seam skeleton and six vertex blooms picked out. The colourbar is clipped
and does \textbf{not} reach the true extremes (min −3.57\%, max
+4.80\%).]{Per-hex area deviation, Mollweide, on a scale clipped at
−0.31\%/+0.26\% --- the p1--p99 band (pattern view). The clip makes the
spatial structure legible: a quiet near-white interior with the octant
seam skeleton and six vertex blooms picked out. The colourbar is clipped
and does \textbf{not} reach the true extremes (min −3.57\%, max
+4.80\%).\footnotemark{}}
\end{figure}
\footnotetext{Generated by
  \protect\texttt{examples/ex0118\_mollweide.py} (L5).}

\begin{figure}
\centering
\includegraphics[width=1\linewidth,height=\textheight,keepaspectratio,alt={Where the ``quasi'' in quasi-authalic lives: every L5 cell classified by signed area deviation, showing one vertex from each of the three vertex classes at a common scale --- (a) North Pole, (b) 0°N 0°E, (c) 0°N 90°E; every hexagon is one L5 cell (≈ 720 km², ≈ 33 km across). All but 244 of the 708,588 cells lie in the white class (within 0.1\% of ideal area). The exceptions cluster at the six octahedral vertices in three pairwise-exact classes: the poles (48 cells beyond 0.1\% each, 12 beyond 1\%, extremes −3.57\%/+4.80\%), the 0°/180° pair (38 and 2, extreme −2.10\%), and the 90°E/W pair (36 and 2, extreme −2.60\%). Within each pair the sorted deviation spectra agree beyond three decimals, so the three classes are structural, not numerical. Morphology follows class: the polar bloom carries an excess (red) core with rays on all four meeting seams; the equatorial blooms have near-tolerance cores with their extremes wholly in deficit (blue), and their rays are meridional only --- the equator seams stay within tolerance throughout.}]{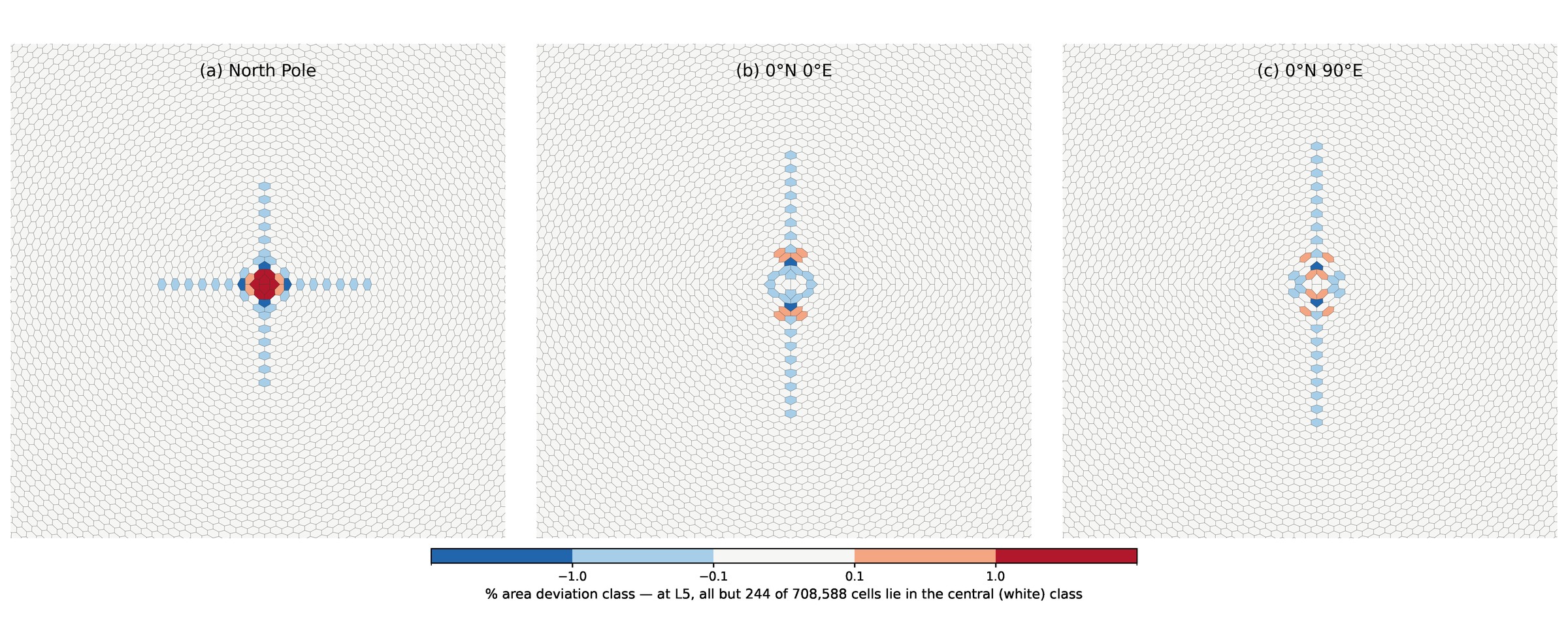}
\caption[Where the ``quasi'' in quasi-authalic lives: every L5 cell
classified by signed area deviation, showing one vertex from each of the
three vertex classes at a common scale --- (a) North Pole, (b) 0°N 0°E,
(c) 0°N 90°E; every hexagon is one L5 cell (≈ 720 km², ≈ 33 km across).
All but 244 of the 708,588 cells lie in the white class (within 0.1\% of
ideal area). The exceptions cluster at the six octahedral vertices in
three pairwise-exact classes: the poles (48 cells beyond 0.1\% each, 12
beyond 1\%, extremes −3.57\%/+4.80\%), the 0°/180° pair (38 and 2,
extreme −2.10\%), and the 90°E/W pair (36 and 2, extreme −2.60\%).
Within each pair the sorted deviation spectra agree beyond three
decimals, so the three classes are structural, not numerical. Morphology
follows class: the polar bloom carries an excess (red) core with rays on
all four meeting seams; the equatorial blooms have near-tolerance cores
with their extremes wholly in deficit (blue), and their rays are
meridional only --- the equator seams stay within tolerance
throughout.]{Where the ``quasi'' in quasi-authalic lives: every L5 cell
classified by signed area deviation, showing one vertex from each of the
three vertex classes at a common scale --- (a) North Pole, (b) 0°N 0°E,
(c) 0°N 90°E; every hexagon is one L5 cell (≈ 720 km², ≈ 33 km across).
All but 244 of the 708,588 cells lie in the white class (within 0.1\% of
ideal area). The exceptions cluster at the six octahedral vertices in
three pairwise-exact classes: the poles (48 cells beyond 0.1\% each, 12
beyond 1\%, extremes −3.57\%/+4.80\%), the 0°/180° pair (38 and 2,
extreme −2.10\%), and the 90°E/W pair (36 and 2, extreme −2.60\%).
Within each pair the sorted deviation spectra agree beyond three
decimals, so the three classes are structural, not numerical. Morphology
follows class: the polar bloom carries an excess (red) core with rays on
all four meeting seams; the equatorial blooms have near-tolerance cores
with their extremes wholly in deficit (blue), and their rays are
meridional only --- the equator seams stay within tolerance
throughout.\footnotemark{}}
\end{figure}
\footnotetext{Generated by
  \protect\texttt{examples/ex0124\_vertex\_census.py} (census table and
  figure; shares the cached geometry of
  \protect\texttt{ex0081w\_warped\_authalics.py}).}

Six irreducible vertex singularities remain at the octahedral poles ---
the geographic N and S poles and the four equatorial points at 0°, 90°,
180°, 270°E. This residual is topological in origin: the Gauss-Bonnet
theorem requires concentrated curvature at the 4-valent octahedral
vertices, and no smooth warp can remove it entirely. The singularities
are fixed and geometrically determined. On WGS84 they are not symmetric:
the geographic poles produce more pronounced residuals than the
equatorial vertices, a consequence of ellipsoidal oblateness
concentrating curvature at the minor-axis tips. The census is sharper
still: the six vertices fall into \emph{three} pairwise-exact classes
--- the poles, the 0°/180° pair, and the 90°E/W pair --- distinct in
count, extreme, sign, and morphology (figure). The polar/equatorial
split follows from oblateness; the split between the two equatorial
pairs cannot (the ellipsoid is rotationally symmetric) and is structural
in origin --- the pairwise-identical spectra rule out numerical noise.
The structure is visible at L0: exactly two root cells meet at each
octahedral vertex (the seed solid, §9), and the axis of their join takes
three orientations against the graticule --- diagonal at the poles,
meridional at 0°/180°, equatorial at 90°E/W. A control run with the warp
disabled confirms the seed: the raw AK realisation separates the same
three classes with the same pairwise exactness, though minutely (class
differences of order 0.005\% inside its ±20\% field); the warp, in
removing the shared bulk deviation, leaves a class-dependent residual
that magnifies the equatorial-pair separation roughly a hundredfold, to
the ≈0.5\% shown.\footnote{Control run:
  \texttt{examples/ex0124\_vertex\_census.py\ -\/-raw} (warp disabled
  via \texttt{b\_oct.no\_warp()}); raw AK leaves only 4,440 of the
  708,588 cells within 0.1\% of ideal area --- the warp brings that to
  all but 244.}

\subsubsection{11c. The Native Space}\label{c.-the-native-space}

After the warp, each x\_cell occupies a well-defined region in b\_oct
--- the octant barycentric space. In b\_oct, the hexagonal cells are
congruent and regular: no cell is larger, smaller, or differently shaped
than any other (modulo the irreducible vertex residuals). b\_oct is the
natural coordinate space for Hex9 computation; it is where the
inequality evaluations of §10a operate and where address digits
correspond to regular geometric subdivisions.

Much of the shape variation that appears when Hex9 cells are rendered in
a conventional map projection --- elongated cells near the poles in
Mercator, apparent distortion in geographic lat/lon --- is a consequence
of that reprojection, not a property of the cells: in their native
b\_oct space the cells are regular by construction, and that added
distortion belongs to the map. This is not to claim the cells are
distortion-free on the ellipsoid itself --- laying an \emph{equal-area}
hexagonal grid on a curved surface costs some bounded shape distortion
(the octahedral and warp distortion of §12, aspect ≈ 1.37), and that
part is intrinsic. The narrower point holds: a conventional projection
\emph{adds} its own distortion on top, and that addition --- not the
grid --- is what a Mercator or lat/lon plot mostly shows.

This is the practical meaning of the claim in §9 that Hex9 is a spatial
reference system rather than merely an indexing scheme: b\_oct is a
legitimate projected coordinate space in its own right, realised by the
AK+Warp projection (§13e). Conventional geographic renderings are
derived from it, not the other way around.

\begin{figure}
\centering
\includegraphics[width=0.95\linewidth,height=\textheight,keepaspectratio,alt={Tissot indicatrices on the AK+Warp b\_oct butterfly net: near-constant-radius circles across the whole net demonstrate equal area, and their near-circularity the low shear the optimal-transport derivation buys (§11b). For the same grid seen through conventional projections --- where the added distortion belongs to the target projection, not the grid --- see the §14c comparison panels.}]{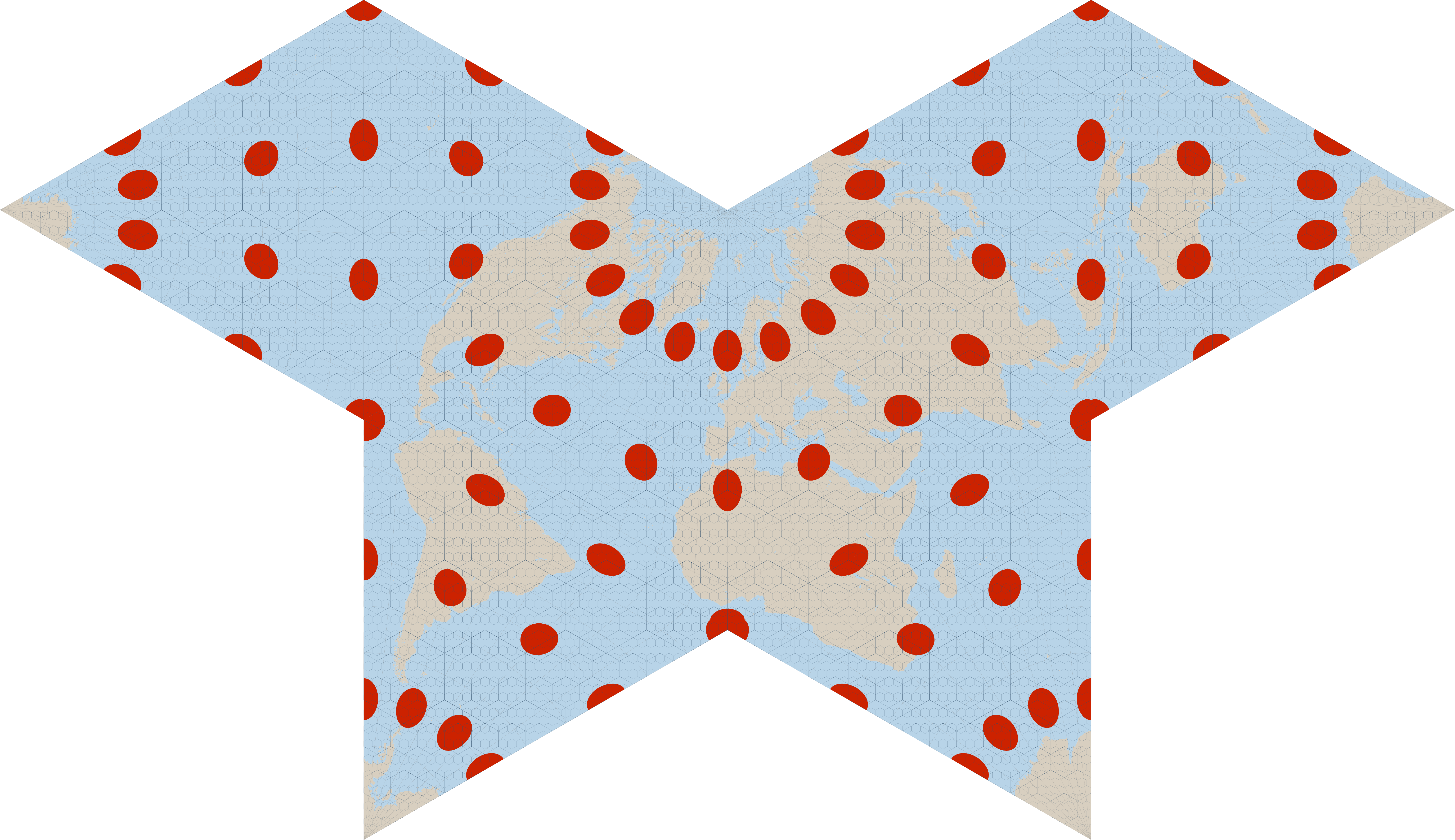}
\caption[Tissot indicatrices on the AK+Warp b\_oct butterfly net:
near-constant-radius circles across the whole net demonstrate equal
area, and their near-circularity the low shear the optimal-transport
derivation buys (§11b). For the same grid seen through conventional
projections --- where the added distortion belongs to the target
projection, not the grid --- see the §14c comparison panels.]{Tissot
indicatrices on the AK+Warp b\_oct butterfly net: near-constant-radius
circles across the whole net demonstrate equal area, and their
near-circularity the low shear the optimal-transport derivation buys
(§11b). For the same grid seen through conventional projections ---
where the added distortion belongs to the target projection, not the
grid --- see the §14c comparison panels.\footnotemark{}}
\end{figure}
\footnotetext{Generated by
  \protect\texttt{examples/ex0121\_tissot\_svg.py}.}

\subsubsection{11d. Projection
Independence}\label{d.-projection-independence}

The AK+Warp combination is the default realisation of the Hex9 grid on
WGS84, chosen because quasi-equal-area cells make point counts and
density estimates geographically comparable. It is not the only valid
realisation.

The combinatorial Hex9 structure is projection-independent. Any
projection from the octahedron to the ellipsoid can be composed with the
addressing scheme to produce a valid Hex9 coordinate system. The Lee
conformal projection (\citeproc{ref-lee1965conformal}{Lee 1965})
produces angle-preserving cells at the cost of non-uniform area; the
Snyder octahedral equal-area projection
(\citeproc{ref-snyder1992equal}{Snyder 1992}) achieves strict det(J) = 1
analytically and offers a closed-form inverse. Both compose cleanly with
the Hex9 hierarchy; the cell identity, digit assignment, and adjacency
structure are unchanged across projection choices.

The warp is also ellipsoid-specific but not ellipsoid-exclusive.
Recomputing the Sinkhorn displacement field against a different
reference body (GRS80, Bessel, a planetary ellipsoid) yields a Hex9
realisation for that body. The addressing hierarchy and all
combinatorial structure carry over unchanged; only the area-correction
data changes. WGS84 and GRS80 are sufficiently close that a single warp
file serves both in practice --- the ellipsoidal difference is
negligible relative to the warp residual.

Projection interchangeability is a consequence of the design philosophy:
the warp corrects for a specific ellipsoid, but the grid it corrects is
defined independently of any ellipsoid. This separation is not an
accident of implementation --- it is the consequence of insisting that
topology, hierarchy, and area correction be treated as distinct concerns
from the outset.

\begin{center}\rule{0.5\linewidth}{0.5pt}\end{center}

\subsection{12. Comparison with Prior
Art}\label{comparison-with-prior-art}

The differences between the established systems --- H3, S2, HEALPix, A5
--- are well documented in the literature; this section makes only the
comparisons that bear on Hex9's claims.

{\def\LTcaptype{none} % do not increment counter
\begin{longtable}[]{@{}
  >{\raggedright\arraybackslash}p{(\linewidth - 10\tabcolsep) * \real{0.1667}}
  >{\raggedright\arraybackslash}p{(\linewidth - 10\tabcolsep) * \real{0.1667}}
  >{\raggedright\arraybackslash}p{(\linewidth - 10\tabcolsep) * \real{0.1667}}
  >{\raggedright\arraybackslash}p{(\linewidth - 10\tabcolsep) * \real{0.1667}}
  >{\raggedright\arraybackslash}p{(\linewidth - 10\tabcolsep) * \real{0.1667}}
  >{\raggedright\arraybackslash}p{(\linewidth - 10\tabcolsep) * \real{0.1667}}@{}}
\toprule\noalign{}
\begin{minipage}[b]{\linewidth}\raggedright
Property
\end{minipage} & \begin{minipage}[b]{\linewidth}\raggedright
H3
\end{minipage} & \begin{minipage}[b]{\linewidth}\raggedright
S2
\end{minipage} & \begin{minipage}[b]{\linewidth}\raggedright
HEALPix
\end{minipage} & \begin{minipage}[b]{\linewidth}\raggedright
A5
\end{minipage} & \begin{minipage}[b]{\linewidth}\raggedright
Hex9
\end{minipage} \\
\midrule\noalign{}
\endhead
\bottomrule\noalign{}
\endlastfoot
Base polyhedron & Icosahedron & Cube & (sphere-native) & Dodecahedron &
Octahedron \\
Cell shape & Hex (+ 12 pentagons) & Quadrilateral & Mixed quad &
Pentagon & Hexagon (no exception cells) \\
Aperture & 7 & 4 & 4 & 5 then 4 & 9 (shifted) \\
Equal area & No & No & Yes (strict) & Yes & Quasi (p99 \textless{}
0.005\%) \\
Strict ancestry at all levels & No & Yes & Yes & --- & Half-hex: yes;
hex: by convention \\
Adjacency isotropy (combinatorial) & Yes & No (\(\sqrt{2}\)) & Yes & No
(elongated) & Yes \\
Dual DGGS/CRS & No & No & No & No & Yes (quasi-CRS) \\
Reference body & Sphere & Sphere & Sphere & Ellipsoid & Any ellipsoid
(per-body warp; WGS84 + sphere trained) \\
\end{longtable}
}

\textbf{Exception cells.} Euler's theorem requires exactly 12 units of
topological defect --- twelve cells short of a full six neighbours ---
in any hexagonal tiling of the sphere, at every refinement level,
independent of resolution. No hexagonal DGGS escapes this; the systems
differ in where the obligation lands. H3 pays it as genuine pentagon
cells: 12 of them at the icosahedral vertices, first-class exceptions
with five children instead of seven, a distinct geometry, and an
\texttt{is\_pentagon()} guard that every correct H3 implementation must
carry --- the price of odd vertex valence, where a 5-valent vertex
cannot sustain the face two-colouring a consistent hexagonal subdivision
depends on (§4). Hex9 introduces no exceptional cell type at all: every
cell is a hexagon. The same obligation appears instead as twelve
five-neighbour hexagons, two at each of the six octahedral vertices,
where two of a hexagon's six sides fold together across the seam to meet
the same opposing hexagon. They carry ordinary addresses, are
constructed by the same d\_cell join rule as every other cell (§10f),
and require no API guard. On the flat net they are plain hexagons;
projected onto a reference ellipsoid one of their six corners falls
collinear, so they render as pentagons while remaining hexagons
combinatorially. In Hex9's native planar net the six vertices lie on the
boundary of the coordinate space, so these singular cells straddle the
edge rather than appearing as interior anomalies.

\textbf{Ancestry.} The strictly nested unit in Hex9 is the half-hexagon
(d\_cell), not the hexagon. The d\_cell hierarchy is a single-parent
tree: a d\_cell address composes left-to-right, prefix-truncation always
yields its unique ancestor, and no edge of the finest-level half-hexagon
tiling crosses a coarser half-hexagon boundary. Hexagons are assembled
from two half-hexagons, and 3 of the 9 child hexagons per level (digits
6--8) straddle a d\_cell boundary and have two valid parents; the
canonical mode-0 convention selects one, and the exact ancestor is
recovered from the address tail (§10b). Under that convention the
canonical parent function is well-defined and deterministic at every
level, and roll-up at the d\_cell level is geometrically exact --- but,
unlike the unconditional quad hierarchies of S2 and HEALPix, hexagon
roll-up requires deriving the canonical ancestor (via the tail) before
truncating, and the split-cell ambiguity can nest: a run of split digits
stays ambiguous until the tail resolves it. H3's aperture-7 subdivision
is exact in the index --- every cell has one deterministic parent, and
parents compose transitively --- but it bears no consistent relation to
geometric containment, and the divergence compounds with depth. Of a
parent's seven children only the central one lies inside it; a
generation further down, boundary grandchildren straddle the parent edge
roughly half-and-half; another generation down, index-descendants appear
that lie entirely outside the ancestor whose address they
extend.\footnote{Demonstrated by \texttt{docs/dggs/dggs\_nesting.py}
  (reference \texttt{h3}, \texttt{s2sphere}, \texttt{a5} and
  \texttt{shapely} libraries): the H3 index hierarchy is single-valued
  and transitive, yet index-descendants become geometrically disjoint
  from their ancestor by the third generation.} The systems trade off
differently rather than one being weaker: Hex9's half-hexagon hierarchy
is exactly nested in area, while its hexagon hierarchy carries a genuine
split-cell ambiguity resolved by the mode-0 convention through the tail;
H3's hierarchy is an exact index laid over grids that never
geometrically refine one another.

The comparison can be made quantitative without privileging any system's
vocabulary. Every hierarchical DGGS offers two routes from a point p to
a coarse cell: bin directly, cell(p, L), or bin fine and truncate,
ancestor\(_L\)(cell(p, L+k)). In fairness, for single-layer binning the
direct route is available and correct in every system, H3 and A5
included; the two routes are distinguished only when layers must be
\emph{related} --- roll-up, drill-down, joining datasets binned at
different resolutions --- which is the work a hierarchy exists to do.
Measured over 10,000 uniformly distributed points, each system through
its own native operations and with no polygon geometry
involved,\footnote{\texttt{docs/dggs/dggs\_commute.py} (N = 10⁴, fixed
  seed); binning at each system's maximum depth (\texttt{-\/-deep})
  reproduces the constant rates and the Hex9 decay law.} the two routes
disagree in H3 for a near-constant ≈ 6.5\% of points at every depth k =
1\ldots6 --- the fixed mismatch between the parent hexagon and the
Gosper-island-like limit shape of its descendant set --- and in A5 for a
constant ≈ 35\% (the rhombus limit of its pentagon refinement). S2's
routes agree everywhere. Hex9's hexagon routes disagree exactly on the
split-cell band, and the rate decays threefold per level --- 16.8\% at k
= 1, 0.07\% at k = 6, following the predicted (1/6)·3\^{}(1−k) ---
because the limit shape of a Hex9 cell's descendant set is the ancestor
itself: the disagreement is a boundary band that thins with depth, not a
shape mismatch that persists. Every disagreeing point is assigned to a
cell that shares the straddling hexagon with its geometric parent --- an
adjacent cell, never a distant one --- and at the d\_cell level the rate
is zero at every depth. In the workload case --- bin once at maximum
resolution, roll up to the analysis layer --- the measured rates are H3
6.5\% (r15 → r5), A5 35.0\% (r30 → r5), and 0.00\% for both S2 (L30 →
L8) and Hex9 (L30 → L5): a constant rate is structural and depth cannot
amortise it, while the Hex9 band decays below any measurable rate
((1/6)·3⁻²⁴ ≈ 10⁻¹³). For Hex9 that deep case is not a favourable corner
but the intended pipeline: a point is encoded once to its full 128-bit
address --- layer 30, a cell some tens of nanometres across --- and
every coarser bin is thereafter derived from the address alone by
truncation and tail canonicalisation, in constant time with no
re-encoding from geometry (§13b). In normal use the system therefore
operates permanently at the measured-zero end of the table.

This is the signature of any hierarchy built by \emph{iterating} a
single-level parent/child relation. H3's aperture-7 step scales by
\(\sqrt{7}\) and rotates by ≈ 19.1°, so each level is turned against the
last and the offset accumulates until descendants leave the ancestor
outright. A5's same-prefix refinement behaves the same way --- under its
own descendant relation its finer cells likewise spill outside the
anchor, and a larger fraction of them than H3's. This is no defect of
A5: it is an equal-area index, and its hierarchy preserves area
continuity rather than geometric containment --- the two cannot be had
together on a curved surface, and A5 makes the opposite trade to S2 and
Hex9. The systems that nest are the ones whose refinement does not
iterate an offset: S2's quadtree, where squares refine onto squares, and
Hex9, whose descendants are not walked down the tree but read directly
from a per-layer contract (§7, §10b) --- so the split hexes, the one
place an iterated step would break, stay bound to their owning ancestor
and never decohere. The commutation rates above are the same distinction
made quantitative: a constant rate is the signature of a limit-shape
mismatch, a geometrically decaying one of an exact limit shape with a
thinning straddle band.

\begin{figure}
\centering
\includegraphics[width=0.9\linewidth,height=\textheight,keepaspectratio,alt={Hierarchy nesting across four DGGS. Each panel takes one anchor cell (heavy outline) and draws its index-descendants three resolutions finer --- by that system's own descendant relation --- coloured by overlap with the anchor: blue lie inside, gold straddle the boundary, red lie entirely outside the ancestor whose address they extend. H3 (aperture-7, rotated each level) and A5 (equal-area, same-prefix children) iterate a single-level step and decohere: index-descendants drift out of the ancestor, A5 the more so. S2's quadtree and Hex9's per-layer contract do not iterate an offset --- S2 nests exactly, and no Hex9 descendant lies outside the anchor (its gold cells are the genuine split hexes, each bound to its owner by the mode-0 convention). Generated by docs/dggs/dggs\_nesting.py at anchor 40°N 3°W; the exact counts are anchor-dependent, the qualitative behaviour is not.}]{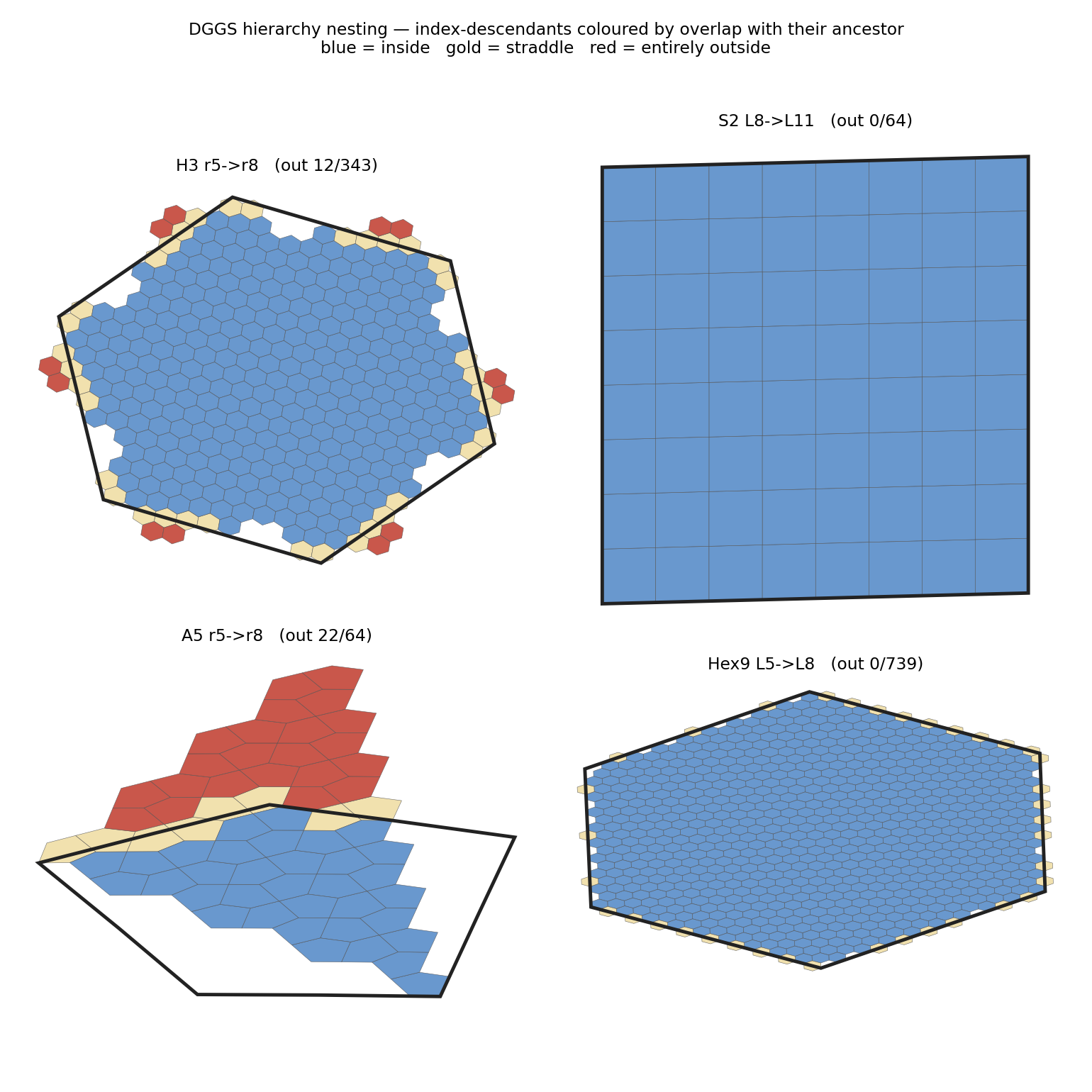}
\caption{Hierarchy nesting across four DGGS. Each panel takes one anchor
cell (heavy outline) and draws its index-descendants three resolutions
finer --- by that system's \emph{own} descendant relation --- coloured
by overlap with the anchor: blue lie inside, gold straddle the boundary,
red lie \textbf{entirely outside} the ancestor whose address they
extend. H3 (aperture-7, rotated each level) and A5 (equal-area,
same-prefix children) \emph{iterate} a single-level step and decohere:
index-descendants drift out of the ancestor, A5 the more so. S2's
quadtree and Hex9's per-layer contract do not iterate an offset --- S2
nests exactly, and no Hex9 descendant lies outside the anchor (its gold
cells are the genuine split hexes, each bound to its owner by the mode-0
convention). Generated by \protect\texttt{docs/dggs/dggs\_nesting.py} at
anchor 40°N 3°W; the exact counts are anchor-dependent, the qualitative
behaviour is not.}
\end{figure}

\textbf{Area.} HEALPix is strictly equal-area but pays in cell shape:
its cells are quadrilaterals of visibly varying geometry. H3 and S2 are
not equal-area; cell areas vary by tens of percent across the globe,
which biases any analysis that compares counts or densities between
regions. Hex9 is quasi-authalic by construction of the warp (§11b): the
residual is bounded, characterised, and confined to the neighbourhoods
of the six vertex singularities.

\textbf{Shape and area, independently surveyed.} An independent survey
of cell geometry --- enclosing-cone aspect ratio and WGS84 geodesic cell
area over N = 5000 uniformly-sampled cells per system (H3, S2, A5, and
Hex9; HEALPix is covered qualitatively under \emph{Area}, below) ---
places the four systems into two area tiers separated by roughly
100×.\footnote{Generated by libhex9 \texttt{tools/dggs\_survey\_hex9.py}
  (N = 5000 cells per system, seed \texttt{0xc0ffee});
  \url{https://github.com/MrBenGriffin/libhex9}. Cell areas are WGS84
  geodesic areas (geographiclib); H3, S2 and A5 are measured through
  their own reference libraries. The surveyed Hex9 polygon is the
  hexagonal cell (the x\_cell of §10a), the same footprint class as an
  H3 hexagon, not the internal half-hexagon.} A5 and Hex9 form an
equal-area tier --- A5 exactly equal-area by construction (measured
coefficient of variation 0.00\%, the residual at the geodesic-area
measurement floor), Hex9 equal-area to a coefficient of variation of
0.10\% through its corrective warp; H3 and S2 do not (12.5\% and
14.5\%), and the gap is a step, not a gradient. The survey reproduces
Hex9's warp characterisation independently --- worst cell ≈ +4.8\%, mean
absolute deviation 0.001\%, all residual pooled at the twelve
octahedral-vertex defects --- on a third-party tool and against the
WGS84 datum, corroborating §11b. On cell shape the order is H3 (aspect
ratio 1.06, roundest) \textless{} S2 (1.24) \textless{} Hex9 (1.37)
\textless{} A5 (2.14): Hex9 reaches the equal-area tier while staying
markedly rounder than the only other equal-area system in the
comparison. Both metrics are resolution-invariant (reproduced to four
decimals five levels coarser), and the dispersion ranking A5 \textless{}
Hex9 \textless{} H3 \textless{} S2 is the same under every statistic
tested (CV, p90, p95, max/min). Hex9 is therefore the better combined
shape-and-area cell among the equal-area systems: A5 attains exact equal
area but with markedly elongated cells (aspect 2.14), whereas Hex9 holds
equal area to a mean absolute deviation of 0.001\% --- its residual
pooled at the twelve octahedral-vertex defects (max/min 1.059) --- while
keeping cells far rounder (aspect 1.37).

\textbf{Aperture.} Hex9's aperture is 9, but the subdivision is not the
centred aperture-9 scheme of the DGGS literature. In a centred scheme
one child sits concentrically at the parent's centroid; in Hex9 the
parent centroid falls on the shared long edge of its two half-hexagons,
so there is no central child and the 9 children form the shifted
half-hexagon arrangement of §8. The shift is not a cost. A centred
aperture-9 grid must descend one triangulation level below its own cells
to describe their boundaries; Hex9's half-hexagon is its own primitive
at every level, and point membership reduces to a fixed set of linear
inequalities (§10a). Comparisons of aperture across systems should note
this distinction.

We coin ``shifted aperture 9'' to be explicit about the offset. Each
parent has exactly nine children --- aperture 9 in the strict sense ---
but they carry a translational shift rather than sitting centred on the
parent. This parallels H3, whose aperture-7 children carry a rotational
offset between successive resolutions yet are still called aperture 7;
we prefer to surface the offset in the name rather than leave it
implicit.

\textbf{Reference body.} H3, S2, and HEALPix are defined on the sphere;
ellipsoidal use requires an auxiliary-latitude transformation with its
own distortion budget. Hex9 is defined against any two-axis reference
ellipsoid directly: the warp is computed from geodesic areas on that
body and recomputed for any other (§11d). Trained warps currently exist
for WGS84 and the sphere; the C implementation (libhex9) is tuned for
WGS84. WGS84 is the default, not a constraint.

\textbf{Adjacency isotropy.} As a graph, a hexagonal tiling gives every
cell six neighbours related by symmetry, with no distinguished direction
--- where a quadrilateral grid distinguishes edge from corner neighbours
(S2's differ by a factor of \(\sqrt{2}\)), forcing distance and
adjacency operations to treat the diagonal as a special case. This is a
\emph{combinatorial} property of the grid graph, not a metric one: once
cells are placed on the ellipsoid, the base projection and the
area-correcting warp make geodesic neighbour distances vary ---
\textbf{no DGGS, Hex9 included, is metrically isotropic on the globe}.
What the hexagonal adjacency buys is algorithmic --- traversal,
\(k\)-ring, and neighbour queries carry no preferred axis. Hex9 and H3
share this adjacency isotropy; S2 does not (the diagonal \(\sqrt{2}\));
A5's pentagonal cells additionally vary in shape (aspect ≈ 2.14).

\textbf{Dual DGGS/CRS role.} This is the row no other system marks
``yes'', and it is the paper's central claim (§9, §10e, Appendix A): a
single Hex9 address is a DGGS zonal identifier when truncated at a level
(OGC Topic 21), and in the limit a function recovers from it a point on
the ellipsoid to arbitrary precision. We state this as the weaker,
defensible claim --- the addressing is \emph{quasi-continuous} (§10e),
not a strict ISO 19111 CRS: the recovery map is discrete-valued and
jumps across a measure-zero set of seams, where continuity in the ISO
sense fails. The distinction from the others is real but one of design,
not exclusive capability: H3, S2, and HEALPix are bounded cell indices
defined over a pre-existing CRS, and as used their identifiers label
cells rather than recover positions. A cleanly-nesting index such as
S2's quadtree could in principle be extended to the same
position-recovery role (§10e claims no uniqueness for the property).
What is intrinsic to Hex9 is that a single address is, by construction,
both the cell identifier and the position-recovery coordinate.

\begin{center}\rule{0.5\linewidth}{0.5pt}\end{center}

\subsection{13. Implementation}\label{implementation}

Hex9 is implemented in \textbf{libhex9}, a C/C++ core library with
several front-ends: a PostGIS extension, a Python accelerator binding,
and a set of Python command-line tools. The earlier Python package
(\texttt{hhg9}) remains as a readable reference implementation and
produced the warp characterisation of §11b. Code artefacts use the
prefix \texttt{h9} (e.g.~\texttt{h9\_encode}); prose uses Hex9
throughout.

\subsubsection{13a. The partition cycle}\label{a.-the-partition-cycle}

The algorithmic heart of encoding is a repeated three-step cycle on
octant coordinates: classify the point into one of the 9 child regions
(the linear inequality evaluation of §10a); subtract the child's origin;
scale by 3. Each iteration emits one address digit and returns the
coordinates to the original numerical scale, so there is no accumulating
floating-point drift; the cycle is O(L) per address, numerically stable,
and exact in integer arithmetic. It is also the CRS limit of §10e made
operational: the offset removal composed with the ×3 scaling is the
inverse of a contraction mapping, and iterating it indefinitely
converges to the encoded point.

Decoding runs the cycle in reverse, with one structural caveat: the
split cells of §10b make the x\_list right-to-left, so the decoder first
reads the tail to fix the terminal mode and then resolves canonical
parentage upward in a single pass.

\subsubsection{13b. Address encoding}\label{b.-address-encoding}

A Hex9 address packs into a single standard 128-bit UUID as 32 nibbles:
30 carry the hierarchy path (one base-9 digit per level, reaching layer
30) and the final two carry the tail (§10b). Because the canonical
mode-0 fold (§10b) gives every cell a single representative, the UUID is
\textbf{self-inverting}: a full address and any coarser \textbf{bin} of
it are the same kind of object --- each decodes, with no companion
field, to its \emph{own} cell's representative point on the ellipsoid.
The two tail forms (a marker nibble distinguishes a bin key from a full
address) differ only in \emph{which} cell is recovered: a bin inverts to
the bin cell's position, a full address to the terminal cell's; neither
recovers the original source point, which binning necessarily discards.
(This supersedes an earlier design in which only the full address was
reversible and a bin was an opaque, non-invertible key.)

The format drops directly into any database, PostGIS column, or API that
accepts a standard UUID; hierarchy traversal is nibble truncation and
spatial binning is prefix comparison, with no decoding step. The
hierarchy reaches layer 30 --- far below any geodetic resolution (a
layer-30 cell is roughly 40 nanometres across) --- so 128 bits is
effectively unlimited precision in practice. The same canonical fit
packs an address into a standard \textbf{64-bit} integer just as readily
--- as the Python library does --- reaching about layer 14: cells about
1.7 m across, sub-metre positional quantisation (centimetre-scale at
layer 17 with denser packing). The 128-bit UUID remains the default
rather than a liability: PostgreSQL and comparable databases treat a
UUID as a first-class, natively indexed type, so its width is no
obstacle; the 64-bit form is offered for pipelines that require an
integer column, not because the narrower key is intrinsically
preferable.

\subsubsection{13c. The core library and
tooling}\label{c.-the-core-library-and-tooling}

The \texttt{libhex9} C/C++ core implements the full pipeline: the AK
base projection with its analytic Jacobian, the authalic warp
(Clough--Tocher interpolant, Newton--Raphson inverse), encoding and
decoding, k-ring neighbour computation, and cell-polygon generation,
with the trained WGS84 warp embedded in the library. A Python
accelerator (\texttt{hex9\_ext}) exposes the core to Python at native
speed with OpenMP batch encoding, and two Python command-line tools wrap
the common workflows: \texttt{h9\_csv} appends a reversible
\texttt{h9\_uuid} column --- and, optionally, canonical bin or label
columns --- to a latitude/longitude CSV, streaming so it handles large
files; \texttt{h9\_choropleth} turns a point CSV into an adaptive Hex9
choropleth as a GeoJSON feature collection, with no PostGIS or QGIS in
the loop. The \texttt{hhg9} Python package remains the readable
reference: it produced the warp characterisation of §11b and the figures
in this paper, with all grid operations vectorised over NumPy arrays.

\subsubsection{13d. PostGIS extension}\label{d.-postgis-extension}

\texttt{libhex9} ships a PostGIS extension (\texttt{postgis\_hex9}) that
exposes the full surface in SQL:
\texttt{h9\_encode}/\texttt{h9\_decode}, \texttt{h9\_bin} and
\texttt{h9\_label}, the cell- and grid-polygon builders
(\texttt{h9\_cell}, \texttt{h9\_grid}), neighbour queries
(\texttt{h9\_kring}, \texttt{h9\_kdisk}, \texttt{h9\_neighbors}),
\texttt{h9\_common\_ancestor} for canonical roll-up, and
\texttt{h9\_adaptive} for population-driven mixed-resolution layers ---
so a Hex9 address is a first-class spatial key inside the database. It
is built on the C/C++ core rather than Python, since most managed
PostGIS environments treat Python as an untrusted language.
(Distribution is handled directly through \texttt{libhex9}: PostGIS is
winding down its third-party extension distribution programme.)

\subsubsection{13e. On not registering a
CRS}\label{e.-on-not-registering-a-crs}

A PROJ/GDAL plugin was prototyped (\texttt{h9\_boct}, a C++ port of the
hierarchy walk and warp inversion), and the round trip matches the
reference to within tens of nanometres. Two distinct objects could in
principle be registered, and they have different standing.

The Hex9 \emph{address system} is not registered, and should not be. The
reason is the one developed in §10e: it is quasi-continuous, not a
strict ISO 19111 CRS, so presenting it through the CRS machinery would
overstate what it is. Its natural standards home is instead the DGGS
track --- OGC Topic 21 and its ISO publication as ISO 19170-1 --- which
exists precisely because discrete hierarchical grids do not fit the CRS
mould. Applications use the core library and its bindings directly,
treating Hex9 as the discrete addressing system it is.

The \emph{AK+Warp projection} (§11) is a different matter. The b\_oct
space it realises is a continuous, real-valued coordinate space over a
geodetic datum, with a smooth forward map and a convergent numerical
inverse --- the ordinary anatomy of an ISO 19111 projected CRS,
untouched by the discreteness argument above. Its unusual features all
have precedents in registered systems: projections without closed-form
inverses invert iteratively throughout PROJ, interrupted layouts are
established practice (Goode homolosine), and grid-backed coordinate
operations are exactly what the NTv2 and geoid-model machinery
distributes. Registering the AK+Warp projection as a projected CRS is
deferred as future work (§15), not declined in principle.

\subsubsection{13f. Inverse projection by guarded
Gauss--Newton}\label{f.-inverse-projection-by-guarded-gaussnewton}

The base-projection inverse --- ellipsoid point to octant face
coordinate --- is the one numerically non-trivial step. \texttt{libhex9}
provides a fast analytic-Jacobian solver for it, root-identical to the
exact beam search in the smooth interior, at one analytic
forward-and-Jacobian evaluation per iteration instead of several
finite-difference evaluations each running an ECEF→geodetic (Bowring)
step. The unknowns are the two face coordinates \((f_x, f_y)\); the
residual \(r = P(f_x, f_y) - E\) is the 3-vector, in ECEF metres,
between the AK forward map \(P\) and the target point \(E\) on the WGS84
ellipsoid. Because the forward map lands exactly on the ellipsoid, the
least-squares root coincides with the true geodetic match --- there is
no auxiliary-latitude (Bowring) step inside the loop. Each iteration
solves the \(2\times2\) normal equations \((J^{\mathsf{T}}J)\,\delta =
-J^{\mathsf{T}} r\) from the \(3\times2\) Jacobian
\(J = \partial P/\partial(f_x,f_y)\), which is fully analytic: the chain
of \(\partial\mathrm{AK}/\partial(u,v,w)\) (tangent-power derivatives
and the ellipsoid-normalising quotient) composed with the affine
\(\partial(u,v,w)/\partial(f_x,f_y)\), and checked against central
differences. A backtracking line search (halving the step until
\(\lVert r
\rVert^2\) decreases) makes it a strictly damped, guarded Gauss--Newton
iteration; a seam/vertex guard and a failure veto route the thin skin
around seams and vertices to the exact beam search, which remains the
reference. The implementation is
\texttt{libhex9/tools/newton\_invert\_aj.h}.

\begin{center}\rule{0.5\linewidth}{0.5pt}\end{center}

\subsection{14. Applications}\label{applications}

\subsubsection{14a. Binning and density
estimation}\label{a.-binning-and-density-estimation}

The primary use case is the one that motivated the quasi-authalic
default: aggregating point observations into cells whose areas are
comparable anywhere on Earth. Encoding is the O(L) partition cycle;
binning is prefix truncation of the UUID --- with the canonical-ancestor
derivation of §10b applied first wherever non-canonical addresses may be
present. Because cell areas are uniform to within the warp residual, raw
counts are densities up to a single global constant: no per-cell area
correction, no latitude adjustment. Grids that are not equal-area (H3,
S2) require explicit area normalisation for the same task, and the
normalisation factor varies by location.

\emph{{[}Figure 15: population density heatmap binned to Hex9 cells, one
country or global. (F15){]}}

Cells need not share a level. Because parentage is exact (§12), a single
layer can mix resolutions --- coarse cells where data is sparse, fine
cells where it is dense --- without T-junctions, seams, or interpolation
between levels, and the mixed layer still rolls up losslessly.
Refinement can therefore be driven by the data itself: subdivide a cell
only while the population it contains stays within a target band,
stopping when the count falls low enough or a depth limit is reached.
Figure 23 shows this for Thimphu, Bhutan, where a population layer
drives adaptive refinement across L5 to L12 --- coarse L5 cells over
empty terrain, refining to L12 along the inhabited valley floors. Each
cell is shaded by population density (count per authalic cell area), so
the fill is directly comparable across levels without area correction
--- a single coherent grid whose cells span seven levels at once. The
layer is still a partition, not an overlay: each location lies in
exactly one cell, because a finer cell is wholly contained in its
coarser ancestor and the refinement keeps only the level its stopping
rule selects --- so cells of different levels tile without overlap.

\begin{figure}
\centering
\includegraphics[width=0.8\linewidth,height=\textheight,keepaspectratio,alt={Thimphu, Bhutan --- population-driven adaptive refinement spanning L5--L12 in one mixed-resolution Hex9 layer; fill is population density per authalic cell area, directly comparable across levels without area correction.}]{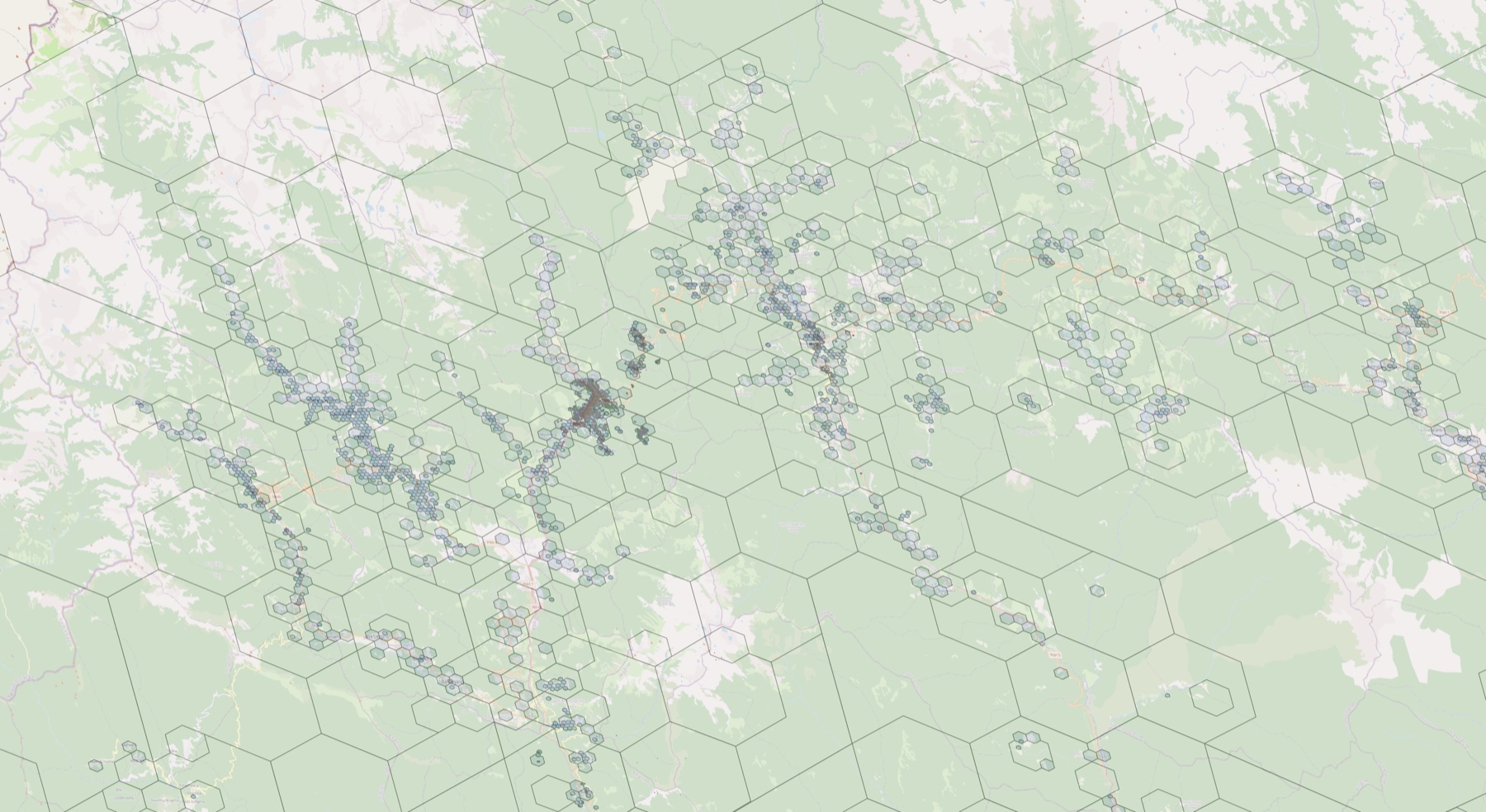}
\caption[Thimphu, Bhutan --- population-driven adaptive refinement
spanning L5--L12 in one mixed-resolution Hex9 layer; fill is population
density per authalic cell area, directly comparable across levels
without area correction.]{Thimphu, Bhutan --- population-driven adaptive
refinement spanning L5--L12 in one mixed-resolution Hex9 layer; fill is
population density per authalic cell area, directly comparable across
levels without area correction.\footnotemark{}}
\end{figure}
\footnotetext{Rendered in QGIS via \protect\texttt{h9\_adaptive()} over
  the Bhutan population layer (\citeproc{ref-hdx_bhutan_pop}{Meta Data
  for Good and Center for International Earth Science Information
  Network (CIESIN), Columbia University 2020}).}

\subsubsection{14b. Spatial joins and multi-resolution
analysis}\label{b.-spatial-joins-and-multi-resolution-analysis}

Two datasets encoded to Hex9 addresses join on shared prefixes: equality
of level-K prefixes is co-location at level K, computed without geometry
--- the full-address truncation whose commutation with direct binning is
measured at 0.00\% over 10⁴ points (L30 → L5, §12). Roll-up from L to
any coarser K is canonical and lossless: every fine cell has exactly one
ancestor at K, so aggregates recombine with no double-counting and no
residue. Its geometric fidelity is governed by the split-cell band of
§12 --- misassignment confined to adjacent cells, decaying threefold per
level of separation, negligible for the deep-to-coarse roll-ups that
dominate practice --- where approximate-parentage systems misassign a
constant fraction at any depth. The same property makes Hex9 addresses
suitable as the spatial component of composite database keys, with
standard B-tree indexes serving range and containment queries that would
otherwise need a spatial index.

\subsubsection{14c. Display and
rendering}\label{c.-display-and-rendering}

The canonical cell polygon is the regular hexagon in b\_oct; everything
else is reprojection (§11c). The equal-area realisation on the ellipsoid
already departs from that regularity by the bounded amount of §12; a
display in a conventional CRS adds more --- cell edges are densified and
reprojected, and the \emph{additional} curvature that appears belongs to
the target projection. Two practical notes. First, hexes straddling the
antimeridian render incorrectly in naive geographic plots (drawn across
the full map width); the defect is the renderer's, not the data's, and a
bounds filter suffices. Second, at the six octahedral vertices the
vertex cells are correct as constructed (§10f) and need no special-case
rendering.

\begin{figure}[ht]
\centering
\begin{minipage}[t]{0.49\linewidth}\centering
\includegraphics[width=\linewidth]{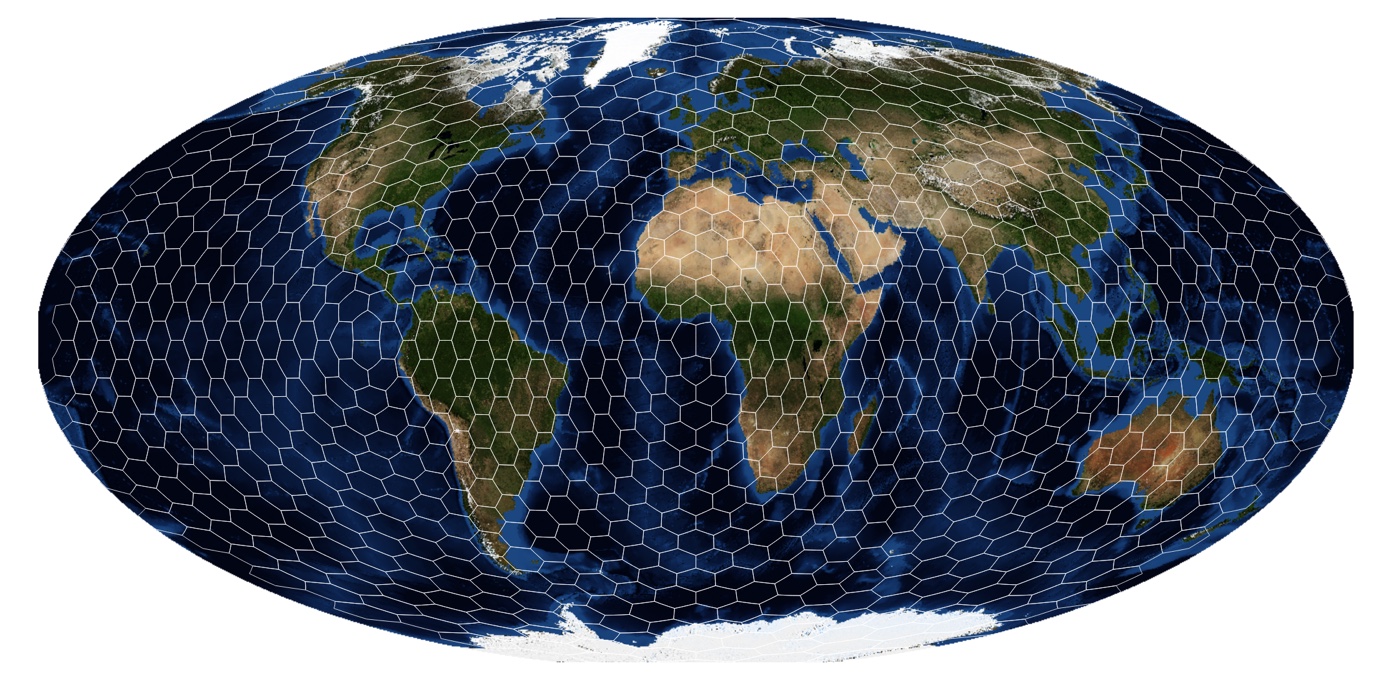}\\[2pt]
{\footnotesize (a) Mollweide (equal-area): areas preserved, shapes shear.}
\end{minipage}\hfill
\begin{minipage}[t]{0.49\linewidth}\centering
\includegraphics[width=\linewidth]{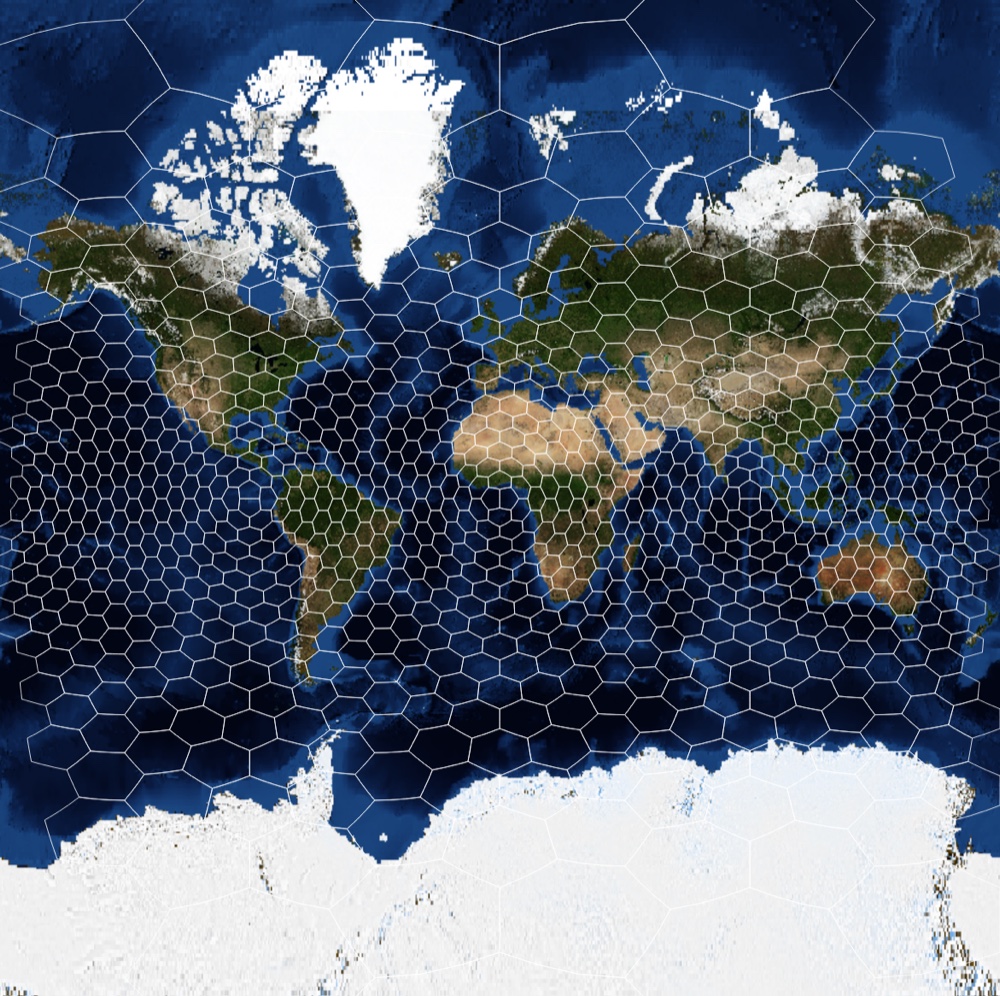}\\[2pt]
{\footnotesize (b) Mercator (conformal): cells balloon toward the poles.}
\end{minipage}
\caption{The same Hex9 layer reprojected --- the distortion a conventional
projection adds belongs to the map, not the cells (\S14c). The native b\_oct
view, in which every cell is a congruent regular hexagon, opens the paper
(Introduction); an equal-area projection (a) keeps cell areas comparable while
shearing their shapes; a conformal projection (b) preserves local angles but
inflates the polar cells without bound (the largest are the octahedral-vertex
cells). Rendered in QGIS. Backdrop: NASA Blue Marble.}
\end{figure}

\subsubsection{14d. Graticule alignment}\label{d.-graticule-alignment}

The octant face spans exactly 90° in latitude and longitude, and 360
carries two factors of 3, so trisection of the degree system is at least
as natural as bisection --- and more useful: bisecting 90° gives 45°,
22.5°, 11.25° (non-standard graticule values), while trisecting gives
30° and 10°, both standard. At 10° spacing the graticule fits the octant
face in 9 clean intervals; L1 hexes span 30°, L2 hexes span 10°. In
arc-minutes the octant (5,400′) trisects cleanly four times (1800′ →
600′ → 200′ → 66.67′) before breaking --- deeper than bisection
achieves. A 10° graticule is therefore the natural companion grid for
Hex9 visualisations: the alignment is a consequence of the shared factor
of 9 between base-360 and the ternary hierarchy, not a design choice in
the projection.

\subsubsection{14e. Other reference
bodies}\label{e.-other-reference-bodies}

The addressing scheme, hierarchy, and interpolation machinery are
ellipsoid-agnostic; only the Sinkhorn warp is computed against a
specific body, at a one-time cost of roughly a day of unattended compute
(§11d). WGS84 and GRS80 share a warp file in practice --- their
difference is negligible against the warp residual. Legacy ellipsoids
(Bessel 1841, Clarke 1866) and planetary bodies with IAU reference
ellipsoids (the Moon, Mars) require only their own warp files and their
own prime-meridian conventions; the mathematical object is unchanged.
Highly irregular bodies (Phobos, small asteroids) are out of scope, as
for every current DGGS.

\begin{center}\rule{0.5\linewidth}{0.5pt}\end{center}

\subsection{15. Conclusion and Future
Work}\label{conclusion-and-future-work}

The argument of this paper runs in one direction: from coherence
requirements to structure, and from structure to location. Orientability
selects the triangle; flat mode transport and vertex closure select even
valence; regularity selects the octahedron; refinement commutativity
selects odd apertures and the minimal aperture 9; the half-hexagon
tiling enumeration --- machine-verified end to end --- leaves exactly
one orientation per chirality. Setting the geometric realisation aside,
precisely two genuine free choices remain in the entire construction:
which chirality, and which bijection assigns the nine digits. Both are
conventions in the same sense as the prime meridian: the grid does not
depend on them, and a variant making the other choice is the same
mathematical object.

Everything the system offers follows from that determinacy. Because no
residual design freedom exists, cell identity is well-defined without
reference to any prior coordinate system, and the address can serve as
both a DGGS cell identifier and, in the limit, a position-recovery
coordinate --- quasi-continuous rather than a strict ISO 19111 CRS
(§10e) --- the dual role developed in §9 and §10e and formalised in
Appendix A. The geometric realisation is a separable concern: the AK
base projection and the optimal-transport warp place the abstract
structure on WGS84 with quasi-uniform areas, and either layer can be
substituted without disturbing the grid.

Future work falls into four groups:

\textbf{Warp refinement.} A strictly authalic variant is achievable in
principle (det J = 1 everywhere) at the cost of unconstrained shear; the
present quasi-authalic compromise is deliberate, but the trade-off curve
--- area residual against shape regularity, via the shape-dampening
parameter --- deserves systematic characterisation. The angular radius
of the elevated-deviation zone around the six vertex singularities is a
one-time, layer-independent measurement not yet made. Deeper-layer warps
(L6+ derived rather than interpolated) would reduce the inter-mode
residual further.

\textbf{Implementation and maintenance.} The core engineering work ---
the \texttt{libhex9} C/C++ library, its PostGIS extension, the Python
accelerator, and the CLI tools (§13) --- is in place; the forward effort
is supporting, hardening, and improving it rather than building new
infrastructure. Warp files for legacy ellipsoids (Bessel 1841, Clarke
1866) and at least one planetary body would demonstrate the generality
claim concretely.

\textbf{Verification.} The R-rotation form of constraint B (§8) remains
available as an independent cross-check of the tiling enumeration. The
C1 continuity of the warp at octant boundaries is asserted from the
construction and should be characterised explicitly.

\textbf{Standardisation and community.} Registration of the Hex9 address
system as a PROJ/GDAL CRS was considered and is not being pursued: it is
better identified as quasi-continuous rather than a strict ISO 19111 CRS
(§10e, §13e), and the standards vehicle appropriate to its DGGS role is
OGC Topic 21 / ISO 19170, not the CRS registry. The AK+Warp projection
is the exception: as a continuous projected coordinate space it is an
ordinary candidate for CRS registration (§13e), and that registration
--- together with a fuller cartographic treatment of the projection,
including systematic distortion comparison against the Snyder and Lee
octahedral projections --- is planned as separate work. Discussions with
PostGIS likewise suggested that it is better to pursue PostGIS
integration as a separate third-party extension rather than look to
integrate into the core; and this thought has been readily accepted. The
path remains open engagement with the discrete-global-grid community
around the existing \texttt{libhex9} implementation, rather than a
formal standards submission. The OGC/ISO terminology mapping of Appendix
B is offered in that spirit --- as a bridge for readers coming from
those standards, not a claim of conformance.

\begin{center}\rule{0.5\linewidth}{0.5pt}\end{center}

\subsection{Appendix A --- The CRS
Limit}\label{appendix-a-the-crs-limit}

\textbf{Claim.} Every infinite Hex9 address sequence σ = (a₀, a₁, a₂,
\ldots) uniquely determines a point on the WGS84 ellipsoid.

\textbf{Setup.} Let Ω ⊂ b\_oct be the closed octant face: a compact
convex triangle of diameter d₀. One step of the Hex9 decode maps Ω into
one of its 9 child cells by

\[p \;\mapsto\; \tfrac{1}{3}\,p + \mathrm{offset}(C_i),\]

a Lipschitz contraction with ratio 1/3. For an address prefix (a₀,
\ldots, a\_L), let S\_L ⊂ Ω be the corresponding cell --- compact and
convex. Then

\[S_0 \supset S_1 \supset S_2 \supset \cdots, \qquad
  \mathrm{diam}(S_L) \;\leq\; d_0 \left(\tfrac{1}{3}\right)^{L} \;\to\; 0.\]

\textbf{Completeness.} Each octant face is a closed bounded subset of ℝ²
and hence a complete metric space; the S\_L are closed subsets of it.

\textbf{Cantor's intersection theorem} (nested compact sets with
diameters tending to zero in a complete metric space) gives

\[\bigcap_{L=0}^{\infty} S_L \;=\; \{p^{*}\},\]

a single point of b\_oct. Equivalently, via Cauchy sequences: choose any
p\_L ∈ S\_L; for L, M \textgreater{} N both p\_L and p\_M lie in S\_N,
so d(p\_L, p\_M) ≤ d₀ (1/3)\^{}N → 0; the sequence is Cauchy, its limit
p* lies in every closed S\_N, and uniqueness follows from diam(S\_L) →
0.

\textbf{Lifting to the ellipsoid.} The composition

\[\text{b\_oct} \xrightarrow{\;\text{AuthalicWarp}^{-1}\;} \text{b\_raw}
  \xrightarrow{\;\text{AK}\;} \text{c\_ell}\]

is continuous: the warp is C1 by the Clough-Tocher construction, and the
AK projection is smooth on the octant interior and at its vertices
(below). Continuous maps preserve limits, so p* maps to a unique point
on the ellipsoid.

\textbf{Continuity at the octahedral vertices.} The octahedral vertices
are the points satisfying \textbar x\textbar{} + \textbar y\textbar{} +
\textbar z\textbar{} = 1 and x² + y² + z² = 1 simultaneously --- exactly
the points with two zero coordinates. There the octahedron and the
sphere coincide, the AK map reduces to the identity followed by axis
scaling to the ellipsoid, and its apparent indeterminacy is a removable
singularity; the Jacobian at these points is the well-defined limit of
the linearisation. No actual singularity exists in the map.

\textbf{Almost-everywhere bijectivity.} The decode map σ ↦ p* is
surjective --- every point is the limit of at least one address chain,
namely its canonical encode --- but not everywhere injective, and the
failure has two distinct sources. The first is point-level and
unavoidable: a point lying on a cell boundary at any level is the limit
of more than one nested chain, one through each closed cell sharing that
boundary --- exactly as 0.5 admits both 0.5000\ldots{} and
0.4999\ldots{} in decimal. The set of such points is the union over all
levels of the cell-edge sets: a countable union of measure-zero sets,
hence measure zero on the ellipsoid. The second is cell-level and
resolved by convention: 3 of the 9 child hexagons per level straddle two
parents (§10b) and would otherwise admit two parent sequences for
\emph{every} point they contain --- a positive-measure ambiguity if left
unresolved. The canonical mode-0 convention selects one parent, making
the canonical address of every cell unique and the encode map p ↦ σ
single-valued everywhere. Away from the measure-zero boundary set,
canonical infinite addresses and ellipsoidal points are in bijection.

\textbf{The dual claim} follows: σ truncated to L digits names a compact
cell of area 510,065,622 km² / (12 · 9\^{}L) on WGS84 (≈ 42.5 million
km² at L0, 719.8 km² at L5) --- a DGGS zone in the sense of OGC Topic
21; σ in the limit names a point, a position recoverable by a function
to arbitrary precision. This is the quasi-continuous coordinate role of
§10e, not a continuous ISO 19111 coordinate system: the recovery map is
well-defined and continuous in the decode direction (infinite address →
point), but the address alphabet is discrete and the encode direction
jumps across the measure-zero seam set just established. The contraction
ratio 1/3 is geometrically determined by the ternary subdivision, so the
convergence rate is explicit: one address digit buys a factor-3
reduction in positional uncertainty.

\begin{center}\rule{0.5\linewidth}{0.5pt}\end{center}

\subsection{Appendix B --- OGC / ISO Terminology
Mapping}\label{appendix-b-ogc-iso-terminology-mapping}

This appendix maps Hex9 vocabulary to the two standards the dual claim
of §9 engages: OGC Abstract Specification Topic 21 (20-040r3, 2021)
(\citeproc{ref-ogc_topic21}{Open Geospatial Consortium 2021}) --- also
published as ISO 19170-1:2021 --- for the DGGS role, and ISO 19111:2019
(\citeproc{ref-iso19111}{International Organization for Standardization
2019}) for the CRS role. The full taxonomy is maintained in the
implementation glossary; the tables here cover the terms a standards
reviewer needs.

Status codes: \textbf{C} --- confirmed (the standard defines the term;
the cited clause was checked against the published text); \textbf{A} ---
analogous (checked against the published text; the correspondence is
real but approximate, as noted); \textbf{---} --- Hex9-specific, with no
standard equivalent. Clause numbers refer to OGC 20-040r3 and to OGC
18-005r4 (the OGC publication of ISO 19111:2019) respectively.

A vocabulary note on Topic 21: the 2021 edition does not simply rename
\emph{cell} to \emph{zone} --- it distinguishes them. A \textbf{zone} is
a ``particular region of space-time'' (4.52), the unit that is
identified; a \textbf{cell} is the ``unit of geometry \ldots{}
associated with a zone'' (4.2). ``While the terms cell and zone are
often used interchangeably, strictly zone is the preferred term. Cell is
entirely appropriate when specifically discussing a zone's geometry or
topology'' (4.2, Note 3). This paper's use of ``cell'' falls within that
licence --- Hex9 cells appear chiefly as geometry; where identification
is meant, the Topic 21 term is \emph{zonal identifier} (4.50).

\subsubsection{Hex9 → OGC Topic 21 (DGGS)}\label{hex9-ogc-topic-21-dggs}

{\def\LTcaptype{none} % do not increment counter
\begin{longtable}[]{@{}
  >{\raggedright\arraybackslash}p{(\linewidth - 4\tabcolsep) * \real{0.3333}}
  >{\raggedright\arraybackslash}p{(\linewidth - 4\tabcolsep) * \real{0.3333}}
  >{\raggedright\arraybackslash}p{(\linewidth - 4\tabcolsep) * \real{0.3333}}@{}}
\toprule\noalign{}
\begin{minipage}[b]{\linewidth}\raggedright
Hex9 term
\end{minipage} & \begin{minipage}[b]{\linewidth}\raggedright
OGC Topic 21 term
\end{minipage} & \begin{minipage}[b]{\linewidth}\raggedright
Status
\end{minipage} \\
\midrule\noalign{}
\endhead
\bottomrule\noalign{}
\endlastfoot
Hex9 (the system) & discrete global grid system (4.13) & C \\
\texttt{x\_grid} at level L & discrete global grid --- one tessellation
per level (4.12) & C \\
\texttt{x\_cell} & zone (4.52); its geometry, a cell (4.2) & C \\
\texttt{x\_adr} / \texttt{x\_list} & zonal identifier (4.50) & C \\
octant (8 seed faces) & face of the base unit polyhedron (4.27) & C \\
12 root \texttt{x\_cells} & initial discrete global grid, refinement
level 0 (4.27, 4.37) & C \\
refinement level L & refinement level (4.37) & C \\
aperture 9 (shifted) & refinement ratio 9 (4.38; Note 3 there records
``aperture'' as earlier DGGS-literature usage) & C \\
binning (\texttt{h9\_bin}) & quantization (4.36) & C \\
parent/child \texttt{x\_cell} & parent cell / child cell (4.33, 4.4) &
C \\
split \texttt{x\_cell} (digits 6--8) & child cell ``overlapped by
multiple parent cells'' (4.4, Note 1) & C \\
\texttt{x\_dig} & --- (ordinal of a child within its parent; no Topic 21
term) & --- \\
AK+Warp realisation & surface model of the Earth (4.27); cf.~Part 1's
Equal Area Earth Reference System (Hex9 is quasi-equal-area, §11b) &
A \\
mode (0/1 parity) & --- (internal face orientation) & --- \\
\texttt{c2} edge label & --- (sub-face structure) & --- \\
\texttt{d\_cell} (half-hexagon) & --- (sub-cell primitive) & --- \\
\end{longtable}
}

\subsubsection{Hex9 → ISO 19111:2019 (referencing by
coordinates)}\label{hex9-iso-191112019-referencing-by-coordinates}

{\def\LTcaptype{none} % do not increment counter
\begin{longtable}[]{@{}
  >{\raggedright\arraybackslash}p{(\linewidth - 4\tabcolsep) * \real{0.3333}}
  >{\raggedright\arraybackslash}p{(\linewidth - 4\tabcolsep) * \real{0.3333}}
  >{\raggedright\arraybackslash}p{(\linewidth - 4\tabcolsep) * \real{0.3333}}@{}}
\toprule\noalign{}
\begin{minipage}[b]{\linewidth}\raggedright
Hex9 term
\end{minipage} & \begin{minipage}[b]{\linewidth}\raggedright
ISO 19111 term
\end{minipage} & \begin{minipage}[b]{\linewidth}\raggedright
Status
\end{minipage} \\
\midrule\noalign{}
\endhead
\bottomrule\noalign{}
\endlastfoot
WGS84 reference ellipsoid & ellipsoid / geodetic reference frame
(3.1.34) & C \\
Prime Meridian anchoring (Axiom 7) & prime meridian (3.1.50) & C \\
\texttt{x\_adr} at a fixed level & spatial reference ``in the form of a
label, code'' (3.1.56); a geographic identifier in the ISO 19112 sense
(via Topic 21 4.50, Note 1) & C \\
\texttt{x\_adr} carried to the limit (§10e, App. A) & approaches the
role of a coordinate --- ``one of a sequence of numbers designating the
position of a point'' (3.1.5) --- quasi-continuous, not a strict ISO
coordinate & A \\
Hex9 as a locating system & coordinate reference system (3.1.9) ---
\emph{quasi}, see §10e & A \\
b\_oct over WGS84, realised by AK+Warp & projected CRS (3.1.51, §9.2.2);
map projection --- ``coordinate conversion from an ellipsoidal
coordinate system to a plane'' (3.1.40) & C \\
encode / decode (point ↔ address) & coordinate conversion (3.1.6) for
the geodetic ↔ b\_oct leg; the address leg assigns a spatial reference
(3.1.56), not a coordinate tuple & A \\
octant 2D plane coordinates & coordinate system / axes (3.1.11) & C \\
\end{longtable}
}

The dual role of §9, §10e, and Appendix A is the bridge between these
two tables: one Hex9 address is a Topic 21 zonal identifier when
truncated at a level and, in the limit, converges to naming a point ---
the role ISO 19111 reserves for coordinates (3.1.5). The standards
themselves mark the boundary the dual claim crosses: Topic 21 notes that
in a DGGS ``locations have dimension greater than one, and so are not
points'' (4.31, Note 2), and that a zonal identifier ``may be a
geographic identifier'' in the ISO 19112 sense (4.50, Note 1) --- a
label, not a coordinate. Hex9's finite-level behaviour sits squarely in
that zonal world; the limit behaviour of Appendix A is what reaches
toward ISO 19111, and we make the weaker, defensible claim for it ---
\emph{quasi-continuous}, by analogy with quasi-authalic --- rather than
asserting a strict ISO 19111 CRS: the address space is discrete and the
point→address map is discontinuous on a measure-zero seam set (§10e), so
continuity in the ISO sense does not hold.

\begin{center}\rule{0.5\linewidth}{0.5pt}\end{center}

\subsection{Figures}\label{figures}

Full figure specifications, production notes, and final captions are
maintained in \texttt{paper\_figures.md}; images live in
\texttt{paper\_figures/}. The index below lists the figures and their
section anchors.

{\def\LTcaptype{none} % do not increment counter
\begin{longtable}[]{@{}
  >{\raggedright\arraybackslash}p{(\linewidth - 4\tabcolsep) * \real{0.3333}}
  >{\raggedright\arraybackslash}p{(\linewidth - 4\tabcolsep) * \real{0.3333}}
  >{\raggedright\arraybackslash}p{(\linewidth - 4\tabcolsep) * \real{0.3333}}@{}}
\toprule\noalign{}
\begin{minipage}[b]{\linewidth}\raggedright
\#
\end{minipage} & \begin{minipage}[b]{\linewidth}\raggedright
Section
\end{minipage} & \begin{minipage}[b]{\linewidth}\raggedright
Working title
\end{minipage} \\
\midrule\noalign{}
\endhead
\bottomrule\noalign{}
\endlastfoot
F1 & §8 / §2 & The 49 tilings, Hex9 pair highlighted \\
F2 & §8 & The two equilateral tilings (L/R chiral pair) \\
F3 & §10.0 & Anatomy of one parent triangle: t / c2 / d / x \\
F4 & §10a & x-layer: x\_dig 0--8 children, high-trit colouring + shared
c-regions \\
F5 & §10a & Classifier (c-layer): 96-slot c\_grid, three band families →
c\_dig \\
F6 & §10b & London / Prime Meridian split-cell example \\
F7 & §10d / §10f & Sibling adjacency: internal vs cross-parent edges \\
F11 & §11 & Pipeline diagram b\_raw → warp → b\_oct → address \\
F12 & §11b & Area-deviation globe, ±1\% capped (magnitude view) \\
F12b & §11b & Area-deviation Mollweide, clipped scale (pattern view:
seam skeleton) \\
F12c & §11b & Vertex census --- the three vertex classes, one panel each
(ex0124) \\
F13 & §11c & Tissot indicatrices on the b\_oct butterfly net (warp
applied) \\
F15 & §14a & Population heatmap binned to Hex9 \\
F16 & §14c & Same layer in b\_oct / Mollweide / Mercator \\
F20 & §2 / §7 / §8 & Wallpaper group of the d\_cell tiling (p31m) \\
F21 & §9 / §10 (graphical abstract) & Seed solid (24 d\_cell facets) +
12 root x\_cells \\
F23 & §14a & Thimphu adaptive refinement (L5--L12), population density
choropleth \\
\end{longtable}
}

\begin{center}\rule{0.5\linewidth}{0.5pt}\end{center}

\subsection*{Acknowledgements}\label{acknowledgements}
\addcontentsline{toc}{subsection}{Acknowledgements}

The octahedral base projection (§11a) is due to Anders Kaseorg
(\citeproc{ref-kaseorg_octahedral}{Kaseorg 2025}). Anthropic's Claude
was used as a drafting and editing assistant; the author alone is
responsible for all claims, derivations, and decisions in this work.

\subsection*{References}\label{references}
\addcontentsline{toc}{subsection}{References}

\protect\phantomsection\label{refs}

\end{document}